\documentclass[reprint,
superscriptaddress,
 amsmath,amssymb,
 aps,
]{revtex4-2}
\usepackage{graphicx}
\usepackage{dcolumn}
\usepackage{bm}
\usepackage{xcolor}
\usepackage{amsmath}
\usepackage[mathlines]{lineno}
\newcommand{\Tr}{\text{Tr}}
\newcommand{\mx}{\text{max}}
\newcommand{\hil}{\mathcal{H}}
\newcommand{\ith}{^{\text{th}}}
\newcommand{\FigScaleFact}{0.25}

\begin{document}

\preprint{APS/123-QED}

\title{Modeling the effects of perturbations and steepest entropy ascent on the time evolution of entanglement}
\author{Cesar  Damian}
  \affiliation{Department of Mechanical Engineering, Universidad de Guanajuato, Salamanca, GTO 36885, Mexico}
 
  \email{cesaredas@vt.edu (sabbatical at Virginia Tech)}
\author{Robert Holladay}
\affiliation{Mechanical Engineering Department, Virginia Tech, Blacksburg, VA 24061}
\author{Adriana Saldana}
\affiliation{Agricultural Engineering Department, Universidad de Guanajuato, Salamanca, GTO 36885, Mexico}
\author{Michael von Spakovsky}%
 \email{vonspako@vt.edu}
\affiliation{Mechanical Engineering Department, Virginia Tech, Blacksburg, VA 24061}%



\date{\today}

\begin{abstract}
This work presents an analysis of the evolution of perturbed Bell diagonal states using the equation of motion of steepest-entropy-ascent quantum thermodynamics (SEAQT), the Lindblad equation, and various measures of loss of entanglement. First, a brief derivation is presented showing that Bell diagonal states are stationary states that are not stable equilibrium states relative to the SEAQT equation of motion, highlighting the need for the development of perturbation methods to study the evolutions of nearby states. This contrasts with the Lindblad equation of motion for which only some of the Bell diagonal states are stationary.  Next, two perturbation methods are presented. The first is a weighted-average method for perturbing bi-partite system states and the second is a general bi-partite method based on a set of unitary operations that are constrained to hold the system energy and system entropy constant. Sets of density operators are randomly generated with each method and the resulting time-varying characteristics of the system's entanglement are analyzed using the SEAQT and Lindblad frameworks.  The findings reveal that the evolutions associated with the constrained perturbations accurately predict the loss of non-locality and align well with the measured concurrence. In addition, using the SEAQT framework, the deep connection between the thermodynamic states of the state evolution of the system and the loss of non-locality is quantitatively demonstrated.
\end{abstract}

\maketitle


\section{Introduction}
Quantum information and computing systems rely on the quantum phenomena of superposition and entanglement to implement useful tasks. In general, two qubits may become entangled if they interact, and certain applications may require that this entanglement persist for some specified amount of time even if the qubits are no longer interacting (such as in the case of quantum teleporting \cite{bennett_teleporting_1993}). Thus, as dissipation occurs throughout a process, it is necessary to understand the effects it has on the entangled state of the system. While various dissipative models predict different effects on the evolution of system state \cite{yu_qubit_2003}, this work focuses on two dynamics of system state evolution. The first is based on the principle of steepest entropy ascent (SEA) as implemented in the SEA quantum thermodynamic (SEAQT) equation of motion \cite{beretta_quantum_1985,beretta_general_1986,beretta_steepest_1987,beretta_steepest_2014,beretta_nonlinear_2006,beretta_nonlinear_2009,beretta_maximum_2010}.  The second is the linear dynamics modeled by the Lindblad equation of motion for open quantum systems. 

A principal aim of this work is to quantitatively demonstrate the deep connection between the thermodynamic states (both non-equilibrium and stable equilibrium) of the state evolution of the system and the loss of non-locality. This is done using the SEAQT framework for which the final stable equilibrium solution found always matches the thermodynamic equilibrium state described by the Gibbs (canonical) distribution. This is generally not the case in the Lindblad framework and occurs only with the correct choice of jump operators. Even when the final state is described by a canonical distribution, the intermediate non-stationary states are not states consistent with the second law of thermodynamics since there is no principle at work in the Lindblad equation that connects the dissipation predicted to this law. The opposite, of course, is true for the SEAQT equation of motion in which the SEA principle is not only consistent with the second law but has as well recently been proposed as a fourth law in its own right \cite{beretta_fourth_2020}.

To compare the predictions of the SEAQT and Lindblad frameworks with experiment, the quantum optics experiment of Liu $et\; al.$ \cite{liu_time-invariant_2016}, which consists of pairs of entangled photons, is used. Experimentally, Liu $et\; al.$ show that the concurrence, which is a measure of the entanglement between the photons, remains mostly unaffected although no error bars are provided. Liu $et\; al.$ furthermore suggest that this is an example of time-invariant entanglement even though the slight variations seen in the data and uncertainties not indicated make their claim questionable, particularly in light of the fact that the state generated by the experiment is probably not a Bell diagonal state due to the noise present \cite{de_vicente_maximally_2024}. More probably, it is a state very close to a Bell diagonal state. Their claim may in part be based on their theoretical predictions of the loss of entanglement between two non-interacting qubits in a Bell diagonal state using the Kraus-Operator-Sum approach. This approach does not assume a specific physical realization for the qubits but does assume that both of the qubits interact with an external global dephasing (decoherence) field, which under this assumption is the cause of the dissipation and the loss of entanglement. Liu $et\; al.$ also show that the maximum expectation value of the Clauser-Horne-Shimony-Holt (CHSH) operator, $\langle{\hat{B}_{CHSH}\rangle}_{\mx}$, which is a measure of non-local quantum correlations, i.e., states whose violation of the inequality cannot be described by local hidden variable models but are still non-signaling, exhibits decay due to the assumed interaction with the dephasing field. Thus, the experiments demonstrate that within the time frame of the experiments, there may be some level of time-invariant entanglement, despite the evolution of the state of the system, while nonlocal quantum correlations experiences sudden death, indicating that stochastic noise correctly approximates the concurrence but not the $\langle{\hat{B}_{CHSH}\rangle}_{\mx}$ value.

In addition, Liu $et\; al.$ \cite{liu_time-invariant_2016} show that some entangled quantum states having values of $\langle{\hat{B}_{CHSH}\rangle}_{\mx} > 2$ undergo an evolution in their dephasing (or decoherence) environment that results in $\langle{\hat{B}_{CHSH}\rangle}_{\mx} < 2$ whereas others exhibit a decay but maintain $\langle{\hat{B}_{CHSH}\rangle}_{\mx} > 2$ and, thus, always produce non-local quantum correlations. This value of 2 is based on the work of Bell \cite{bell_einstein_1964} and Clauser $et \; al.$ \cite{clauser_proposed_1969}, who demonstrate that Local-Hidden-Variable theories can only produce entanglements with a maximum value of 2 for $\langle{\hat{B}_{CHSH}\rangle}$. Cirel'son \cite{cirelson_quantum_1980} later shows that when accounting for quantum phenomena, a value as high as $2\sqrt{2}$ can be obtained. For either value, the implication of a drop in value of $\langle{\hat{B}_{CHSH}\rangle}_{\mx}$ below 2 (or $2\sqrt{2}$) is that entanglement resulting from measurements of a quantum system can no longer be attributed to non-local quantum correlations of the quantum system, as it is possible for entangled systems to exhibit values of $\langle{\hat{B}_{CHSH}\rangle}_{\mx}$ lower than 2 (or $2\sqrt{2}$). This then is what is referred to as the sudden death of non-locality.

Theoretically, the loss of non-locality can be translated to the effect of dephasing due to an external dephasing field as is done by Liu $et\; al.$ \cite{liu_time-invariant_2016} with their Kraus-Operator-Sum model or as is done in the quantum-open-system model or so-called Lindblad model. It can as well be taken into account via intrinsic dissipation due to internal irreversibiliites as in the SEAQT framework. The present work quantifies the effects of the SEA principle as well as that of the Lindblad equation on the dynamics of an entangled system and, consequently, on how the entanglement as measured by the concurrence and $\langle{\hat{B}_{CHSH}\rangle}_{\mx}$ changes with system state. The results are compared to both the theoretical and experimental results of Liu $et\; al.$ \cite{liu_time-invariant_2016}. However, since Bell diagonal states are non-dissipative in the SEAQT framework, two different perturbation methods are used to produce initial states for the SEAQT equation of motion that are physically similar to the original Bell diagonal state: a weighted-average method for perturbing bipartite system states and a general bi-partite method \cite{montanez-barrera_method_2022} based on a set of unitary operations that are constrained to hold the system energy and entropy to be that of the original Bell diagonal state. The same initial states are used with the Lindblad equation even though generally Bell diagonal states are only non-dissipative in the quantum open-system framework under certain conditions. As will be seen, the type of perturbation method used to generate the initial perturbed state determines whether or not the dynamic SEAQT and Lindblad predictions are able to simultaneously match the experimental values for both the concurrence and $\langle{\hat{B}_{CHSH}\rangle}_{\mx}$. 

To accomplish all of this, Section \ref{Sec:Model} first presents the measures used to quantify system entanglement. This is followed by a description and discussion of the SEAQT and Lindblad equations and the Hamiltonian used to model the experiment. The behavior of the perturbed Bell diagonal states relative to the two equations of motion is also discussed, and because Bell diagonal states are non-dissipative in the SEAQT framework, the two types of perturbation methods used here to perturb the non-dissipative initial Bell diagonal state to a dissipative one are presented. The resulting dynamics relative to the SEAQT and Lindblad equations of motion are presented in Section \ref{Sec:Results}. Finally, Section \ref{Sec:Conclusions} provides the primary conclusions of this work.

\section{Model Development}
\label{Sec:Model}

To model the evolution of entanglement due to the principle of steepest entropy ascent, three entanglement measures and the SEAQT equation of motion are first reviewed. The entanglement measures are reviewed in Section \ref{Sec:Entanglement} and are used to examine how the entanglement changes throughout the system's evolution. Next, Section \ref{Sec:SEAQT} presents the SEAQT equation of motion for a general quantum system and how it is used to predict the dissipation and associated change in entanglement characteristics. The class of states of interest in this work are known as Bell-diagonal states and turn out to be non-dissipative relative to the SEAQT equation of motion (a brief proof of which is also presented in Section \ref{Sec:Results}). Thus, to examine the dynamics of dissipation associated with SEAQT, it is necessary to generate dissipative, perturbed states from a initial Bell diagonal state. Section \ref{Sec:Lindblad} then presents the Lindblad formulation where Lindblad operators are proposed to represent a pure dephasing and amplitude damping envixronment independent of the state of the system and interpreted as a bath of photons interacting with the system. The two methods for creating the required perturbed states are presented in Section \ref{Sec:Perturb} and common features relating the perturbed state to that of the initial Bell diagonal state are highlighted. 

\subsection{Entanglement Measures}
\label{Sec:Entanglement}

To analyze the evolution of the entanglement within the SEAQT framework, two different metrics are used: i) the concurrence, $E(\hat{\rho})$, and ii) the maximum expectation value of the CHSH operator, $\langle{\hat{B}_{CHSH}\rangle}$, were $\hat{\rho}$ is the density operator. Many others exist (for a rather comprehensive review see \cite{horodecki_quantum_2009}). The relative entropy $D (\hat{\rho} || \hat{\rho}_0 )$, which is related to the difference between two states, is also explored to measure the distance between the initial Bell diagonal state, $\hat{\rho}_0$, and the perturbed state, $\hat{\rho}$.

The concurrence, $E(\hat{\rho})$, proposed by Hill and Wooters \cite{hill_entanglement_1997} and Wooters \cite{wootters_entanglement_1998} is intended to be a faithful representation of the entanglement and is computed as 
\begin{equation}
E(\hat{\rho}) = \max(0,\lambda_1 - \lambda_2 - \lambda_3 - \lambda_4)
\end{equation}
where the $\lambda_i$ are the eigenvalues of the operator $\hat{R}$ in decreasing order as $i$ increases and $\hat{R}$ is determined using the relation
\begin{equation}
\hat{R} = \sqrt{\sqrt{\hat{\rho}}\hat{\tilde{\rho}}\sqrt{\hat{\rho}}}
\end{equation}
Here the matrix $\hat{\tilde{\rho}}$ is computed as
\begin{equation}
\hat{\tilde{\rho}} = \left(\hat{\sigma}_Y\otimes\hat{\sigma}_Y\right)
\hat{\rho}^*
\left(\hat{\sigma}_Y\otimes\hat{\sigma}_Y\right)
\end{equation}
where $\hat{\rho}^*$ is the complex conjugate of $\hat{\rho}$ and $\hat{\sigma}_Y$ is the Pauli-y matrix. 

The second common entanglement measure used here is the maximum expectation value of the CHSH operator, $\langle{\hat{B}_{CHSH} \rangle}_{\mx}$, which is expressed as
\begin{equation}
\langle{\hat{B}_{CHSH}\rangle}_{\mx} = 2\sqrt{h_i + h_j}
\end{equation}
where $h_i$ and $h_j$ are the two largest eigenvalues of the matrix $T_{\hat{\rho}}^T T_{\hat{\rho}}$ with $T_{\hat{\rho}}^T$ the transpose of $T_{\hat{\rho}}$  and the elements of $T_{\hat{\rho}}$ given by $t_{ij} = \Tr(\hat{\rho}\left(\hat{\sigma}_i\otimes\hat{\sigma}_j\right))$. Here again the $\hat{\sigma}_i$ and $\hat{\sigma}_j$ are the Pauli  matrices used to construct $T_{\hat{\rho}}$.

Now, for a composite system in a dissipative state consisting of subsystems $A$ and $B$, the state of each subsystem in the SEAQT framework evolves to a unique local equilibrium (canonical) state given by $\hat{\rho}_{\text{eq},A(B)} = e^{-\beta_{A(B)}\hat{H}_{A(B)}}/Z_{A(B)}$ with each canonical state in mutual stable equilibrium with the other. In this expression, $\hat{H}$ is the Hamiltonian, $Z$ is the partition function and $\beta$ is a Lagrange multiplier inversely proportional to the temperature of the state. In the Lindblad formulation, the subsystem states evolve towards stationary states that are characterized by a time-invariant density matrix under the dynamics governed by the Lindblad equation. However, these stationary states (i.e., steady states) are not equilibrium (or canonical) states since under the SEAQT dynamics they are dissipative and continue to evolve toward stable equilibrium. 

Finally, to analyze the effects of a perturbation on the state of a system, the quantum relative entropy, which measures the distinguishability between two states, is used. It is given by
\begin{equation}
D( \hat \rho || \hat \rho_0 ) =  \Tr \hat \rho \ln \hat \rho - \Tr \hat \rho \ln \hat \rho_0 \,.
\end{equation}
When there is no perturbation, the relative entropy is zero. Otherwise, it is positive, sometimes decreasing and then increasing or always increasing as the state moves away from the Bell diagonal state employed as our initial state.\\

\subsection{SEAQT Equation of Motion for a General Quantum System}
\label{Sec:SEAQT}
Because entanglement is inherently a feature of composite system states, analyzing the change in entanglement requires analyzing the dynamics of the entire composite system \cite{beretta_quantum_1985,beretta_general_1986,beretta_steepest_1987,beretta_steepest_2014,beretta_nonlinear_2006,beretta_nonlinear_2009,beretta_maximum_2010}. The associated Hilbert space for an isolated composite is system  $\hil = \bigotimes_{J = 1}^M \hil_J$ where $M$ is the number of subsystems in the composite system. Thus, to understand the effects that the dissipation associated with the SEA principle has on the time evolution of the entanglement, the SEAQT equation of motion for a general quantum system is required and is expressed as
\begin{equation}
\frac{d\hat{\rho}}{dt} = -\frac{i}{\hbar}\left[ \hat{H}, \hat{\rho} \right] - \frac{\hat{D}\left(\hat{\rho}\right)}{Dt}\label{EOM9}
\end{equation}
where the dissipation operator $\hat{D}\left(\hat{\rho}\right)/Dt$  for a composite of $M$ subsystems is 
\begin{equation}
\frac{\hat{D}\left(\hat{\rho}\right)}{Dt} = \sum_{J = 1}^{M} \frac{1}{\tau_J}\hat{D}_J\otimes\hat{\rho}_{\bar{J}}
\end{equation}
Here the partial trace of the composite density operator onto the $J\ith$ subsystem is given by $\hat{\rho}_{J} = \Tr_{\bar{J}}(\hat{\rho})$ where $\bar{J}$ represents the composite of all subsystems $\neq J$. Likewise, the density operator for the composite of all subsystems except the $J\ith$ can be computed as $\hat{\rho}_{\bar{J}} = \Tr_{J}(\hat{\rho})$. The dissipation operator for the $J\ith$ subsystem, $\hat{D}_J$, is then given by
\begin{equation} \label{eq:disp}
\hat{D}_J= \frac{1}{2}\left(\hat{\rho}_J\tilde{D}_J + \left(\hat{\rho}_J\tilde{D}_J\right)^\dagger\right)
\end{equation}
where $\tilde{D}_J$ represents a ratio of determinants for the $J\ith$ subsystem that ensures that the expectation values associated with the generators of the motion, $\hat{R}_{iJ}$, of the $J\ith$ subsystem remain constant. $\tilde{D}_J$ is then computed as
\begin{widetext}
\begin{equation}
\tilde{D}_J = \sqrt{\hat{\rho}_J}
\frac{
	\left|\begin{matrix}
	\hat{B}_J\ln(\hat{\rho}_J) & \hat{R}_{1J} & \hat{R}_{2J} & \cdots & \hat{R}_{N_JJ} \\
	\left(\hat{R}_{1J},\hat{B}_J\ln(\hat{\rho}_J)\right) & \left(\hat{R}_{1J},\hat{R}_{1J}\right) & \left(\hat{R}_{1J},\hat{R}_{2J}\right) & \cdots & \left(\hat{R}_{1J},\hat{R}_{N_JJ}\right)
	\\
	\left(\hat{R}_{2J},\hat{B}\ln(\hat{\rho})\right) & \left(\hat{R}_{2J},\hat{R}_{1J}\right) & \left(\hat{R}_{2J},\hat{R}_{2J}\right) & \cdots & \left(\hat{R}_{2J},\hat{R}_{N_JJ}\right)
	\\
	\vdots & \vdots & \vdots & \ddots & \vdots 
	\\
	\left(\hat{R}_{N_J)J},\hat{B}\ln(\hat{\rho})\right) & \left(\hat{R}_{N_JJ},\hat{R}_{1J}\right) & \left(\hat{R}_{N_JJ},\hat{R}_{2J}\right) & \cdots & \left(\hat{R}_{N_JJ},\hat{R}_{N_JJ}\right)
	\end{matrix}\right|
}
{
	\Gamma_J
}\label{DisopJ9}
\end{equation}
where $\hat{B}_J$ is the idempotent operator defined in \cite{beretta_nonlinear_2009}, the subscript $N_J$ is the number of generators of the motion for subsystem $J$, and the Gram determinant $\Gamma_J$ is computed as
\begin{equation}
\Gamma_J = 
\left|\begin{matrix} \left(\hat{R}_{1J},\hat{R}_{1J}\right) & \left(\hat{R}_{1J},\hat{R}_{2J}\right) & \cdots & \left(\hat{R}_{1J},\hat{R}_{N_JJ}\right)
\\ \left(\hat{R}_{2J},\hat{R}_{1J}\right) & \left(\hat{R}_{2J},\hat{R}_{2J}\right) & \cdots & \left(\hat{R}_{2J},\hat{R}_{N_JJ}\right)
\\
\vdots & \vdots & \ddots & \vdots 
\\ \left(\hat{R}_{N_JJ},\hat{R}_{1J}\right) & \left(\hat{R}_{N_JJ},\hat{R}_{2J}\right) & \cdots & \left(\hat{R}_{N_JJ},\hat{R}_{N_JJ}\right)
\end{matrix}\right| \label{DisopJ10}
\end{equation}
\end{widetext}
Each inner product $\left(\hat{F}_J,\hat{G}_J\right)$ is defined as
\begin{equation} \label{eq:inner}
\left(\hat{F}_J,
\hat{G}_J\right) = \frac{1}{2}\Tr_{J}\left( | \hat{\rho}_J | \{ \hat{F}_J , \hat{G}_J \} \right)
\end{equation}
where the anti-commutator $\{ \hat{F}_J , \hat{G}_J \} = \hat{F}_J\hat{G}_J + \hat{G}_J , \hat{F}_J$ and the so-called locally perceived operator $\hat{F}_J$ is computed as
\begin{equation}
\hat{F}_J = \Tr_{\bar{J}}\left(\left(\hat{I}_J\otimes\hat{\rho}_{\bar{J}}\right)\hat{F}\right)
\end{equation}
$\hat{G}_J$ is defined in a similar fashion. In these equations, $\hat{I}$ is the identity operator, $\hat{F}$ and $\hat{G}$ are self-joint operators on Hilbert space $\mathcal{H}$, and $\hat{F}_J$ and $\hat{G}_J$ are operators on $\mathcal{H}_J$ and are, in general, non-self-joint.

To take into account energy exchange with a thermal reservoir (environment), each subsystem $J$ consists of a composite of a system $S$ and reservoir $R$ between which the energy exchange takes place. The energy exchange accounts for the amplitude damping of $S$. The generators of the motion for this composite of $S$ and $R$ of subsystem $J$ then become $\hat R_{1J} = \hat I_{S_J} \oplus \hat 0_R$, $R_{2J} = \hat 0_{S_J} \oplus I_R$, and $R_{3J} = \hat  H_{S_J} \oplus \hat H_R = \hat{H}_J$ where $\hat{0}$ is the null operator and $\hat{H}$ the Hamiltonian. The first of these ensures conservation of probabilities (i.e., the unit trace of $\hat{\rho}_J$) for $S$ and the second ensures that at stable equilibrium the density matrix, $\hat{\rho}_J$, coincides with that of a Gibbs (canonical) state with inverse temperature equal to the inverse temperature, $\beta_R$, of the reservoir $R$. The third generator of the motion ensures conservation of energy for subsystem $J$, i.e., the composite of $S$ and $R$. Note that in the special case of interactions with a thermal reservoir, the Hilbert space of $R$ is considered to be orthogonal to the Hilbert space of $S$, such that it factorizes as $\mathcal{H}_{S_J} \oplus \mathcal{H}_R$ for each subsystem \cite{holladay_steepest-entropy-ascent_2019,li_steepest-entropy-ascent_2016}. In this manner, the spectral properties of $S$ and $R$ are independent.
Eqs. (\ref{DisopJ9}) and (\ref{DisopJ10}) are now written as
\begin{widetext}
 \begin{equation} 
\tilde{D}_J = 
\sqrt{\hat{\rho}_J}
\frac{
	\left|\begin{matrix}
	\left(\hat{B}_J\ln(\hat{\rho}_J)\right) & \left(\hat{I}_{S_J}\oplus \hat{0}_R\right) & \left(\hat{0}_{S_J} \oplus \hat{I}_R\right) & \left(\hat{H}_{S_J}\oplus\hat{H}_R\right)  \\
	\left(\hat{I}_{S_J} \oplus \hat{0}_R,\hat{B}_J\ln(\hat{\rho}_J)\right) & \left(\hat{I}_{S_J} \oplus \hat{0}_R, \hat{I}_{S_J} \oplus \hat{0}_R \right) 
 & \left(\hat{I}_{S_J} \oplus \hat{0}_R, \hat{0}_{S_J} \oplus \hat{I}_R \right) & \left(\hat{I}_{S_J} \oplus \hat{0}_R, \hat{H}_{S_J} \oplus \hat{H}_R\right) 
	\\
	\left(\hat{H}_{S_J} \oplus \hat{0}_R,\hat{B}_J\ln(\hat{\rho}_J)\right) & \left(\hat{0}_{S_J} \oplus \hat{I}_R,\hat{I}_{S_J} \oplus \hat{0}_R \right) & \left(\hat{0}_{S_J} \oplus \hat{I}_R, \hat{0}_{S_J} \oplus \hat{I}_R \right) & \left(\hat{0}_{S_J} \oplus \hat{I}_R, \hat{H}_{S_J} \oplus \hat{H}_R\right) \\
 \left(\hat{H}_{S_J} \oplus \hat{H}_R,\hat{B}\ln(\hat{\rho})\right)  & \left(\hat{H}_{S_J} \oplus \hat{H}_R , \hat{I}_S \oplus \hat{0}_R \right) & \left(\hat{H}_{S_J} \oplus \hat{H}_R , \hat{0}_S \oplus \hat{I}_R \right) & \left(\hat{H}_{S_J} \oplus \hat{H}_R, \hat{H}_S \oplus \hat{H}_R \right)
	\end{matrix}\right|
}
{
	\left|\begin{matrix}
	 \left(\hat{I}_{S_J} \oplus \hat{0}_R,\hat{I}_{S_J} \oplus \hat{0}_R \right) 
 & \left(\hat{I}_{S_J} \oplus \hat{0}_R , \hat{0}_{S_J} \oplus \hat{I}_R \right)  & \left(\hat{I}_{S_J} \oplus \hat{0}_R, \hat{H}_{S_J} \oplus \hat{H}_R\right) 
	\\
	\left(\hat{0}_{S_J} \oplus \hat{I}_R,\hat{I}_S \oplus \hat{0}_R \right) & \left(\hat{0}_{S_J} \oplus \hat{I}_R, \hat{0}_{S_J} \oplus \hat{I}_R \right) & \left(\hat{0}_{S_J} \oplus \hat{I}_R, \hat{H}_{S_J} \oplus \hat{H}_R\right) \\
\left(\hat{H}_{S_J} \oplus \hat{H}_R , \hat{I}_{S_J} \oplus \hat{0}_R \right) & \left(\hat{H}_{S_J} \oplus \hat{H}_R , \hat{0}_{S_J} \oplus \hat{I}_R \right) & \left(\hat{H}_{S_J} \oplus \hat{H}_R, \hat{H}_{S_J} \oplus \hat{H}_R \right)
	\end{matrix}\right|
} \label{DisopJ13}
\end{equation}

\end{widetext}
which can be reduced by determining the inner products and identifying the various entropy and energy expectation values, i.e., $\langle s \rangle_{K}$, $\langle e \rangle_K$, $\langle e^2 \rangle_K$, and $\langle es \rangle_K$ where $K = S_J,R$. Thus, Eq. (\ref{DisopJ13}) takes the form
\begin{equation}
    \tilde D_J = \sqrt{\hat \rho_{S_J}} 
     \frac{
\left| \begin{matrix}
\hat{B}_J\ln(\hat{\rho}_J) & \hat{I}_{S_J}\oplus \hat{0}_R  & \hat 0_{S_J} \oplus \hat{I}_R & \hat{H}_{S_J}\oplus\hat{H}_{R} \\
\langle s \rangle_{S_J} & P_{S_J} & 0 & \langle e \rangle_{S_J} \\
\langle s \rangle_R & 0 & P_{R} & \langle e \rangle_R \\
\sum_K \langle es \rangle_K  & \langle e \rangle_{S_J} & \langle e \rangle_R & \sum_K \langle e^2 \rangle_K 
\end{matrix} \right|
    }
    {
    \left|
\begin{matrix}
P_{S_J} & 0 & \langle e \rangle_{S_J} \\
0 & P_{R} & \langle e \rangle_R \\
\langle e \rangle_{S_J} & \langle e \rangle_R & \sum_K \langle e^2 \rangle_K 
\end{matrix}\right|
    }\label{Eq:DtildeJ}
\end{equation}
Expanding the determinants and using the fact that the ratio of the inner products $\epsilon = \frac{P_{S_J}}{P_R}$ is very small due to the large size of $R$, Eq. (\ref{Eq:DtildeJ}) can be reduced to its final form of
\begin{align}
\hat{D}_J = \hat{\rho}_J \ln \hat{\rho}_J - \beta_R \langle f \rangle_J \hat{\rho}_J +\frac{1}{2} \beta_R \{ \hat{H}_J , \hat{\rho}_J \}\label{Eq:DtildeJfinal}
\end{align}
where $\langle f \rangle_J = \text{Tr}\, \left( \hat{\rho}_J \hat{H}_J\right)+\beta_R^{-1} \text{Tr}\, \left( \hat{\rho}_J \hat{B}_J \ln \hat{\rho}_J \right)$ is the non-equilibrium expectation value of the free energy and 
\begin{equation}
    \beta_R = \frac{\langle e s \rangle_R - \langle e \rangle_R \langle s \rangle_R}{\langle e^2 \rangle_R - \langle e  \rangle^2_R} +\mathcal{O} \left( \epsilon \right)
\end{equation}
matches the equilibrium temperature of a reservoir in a Gibbs state. A detailed derivation of this equation can be found in Chapter 2 of \cite{holladay_steepest-entropy-ascent_2019}.

Absent a reservoir, Eqs.~(\ref{DisopJ9}) and (\ref{DisopJ10}) when substituted with the $\hat{\rho}_J$, $\hat{I}_J$, and $\hat{H}_J$ operators result in a 3$\times$3 determinant in the numerator and a $2\times2$ Gram determinant. The equation of motion with this formulation has been shown to effectively simulate pure dephasing \cite{montanez-barrera_loss--entanglement_2020}, which in the SEAQT framework is due to the dissipation of the system that occurs as a result of internal irreversibilities only. In contrast, the formulation given in Eq. (\ref{Eq:DtildeJfinal}) accounts for both dephasing and the amplitude damping due to energy interactions with an environment and has been shown to effectively simulate both \cite{montanez-barrera_method_2022}. It is this equation, which is used in the SEAQT equation of motion to predict the evolution results presented here. 

Now, for a composite system consisting of subsystems $J=A,B$, the Hamiltonian can be written as 
\begin{eqnarray}
    \hat H &=& \left( \hat H_{S_A} 
\oplus \hat I_R + \hat I_{S_A} \oplus \hat H_{R} \right) \otimes (\hat{I}_R \oplus \hat{I}_{S_B} ) \nonumber \\
&\;& \; +( \hat{I}_{S_A}\oplus\hat{I}_R ) \otimes \left( \hat H_{S_B} 
\oplus \hat I_R + \hat I_{S_B} \oplus \hat H_{R} \right)
\end{eqnarray}
where the $\hat{I}_{{S_A}({S_B})}$ are the identity operators in Hilbert spaces $\hil_{{S_A}({S_B})}$ and the composite system identity operator can be computed as $\hat{I} = \left(\hat{I}_{S_A}\oplus\hat{I}_R\right)\otimes\left(\hat{I}_{S_B}\oplus\hat{I}_R\right)$. While it is well known that two systems must interact to become entangled, this work examines the evolution of two systems that have already interacted to become entangled and are no longer interacting, implying that the interaction term $\hat{V}_{AB} = 0$. Thus, with this model, the subsystems no longer interact to change the entanglement characteristics through the reversible dynamics, and any change in the entanglement characteristics of the composite system are, therefore, solely attributable to the internal dissipation and the energy interaction with $R$.

The dynamics of the SEAQT equation of motion are examined below for a specific class of entangled states, known as Bell diagonal states, to better understand how the entanglement changes. However, a discussion of the Lindblad equation of motion is provided first in the next section.

\subsection{Lindblad Equation of Motion}
\label{Sec:Lindblad}
The theory of open quantum systems offers a way to determine the evolution of a system subject to external perturbations by an environment. The equation of motion for this framework is the Lindblad equation. To compare the loss of non-locality as well as the loss of concurrence obtained by SEAQT formulation with that predicted by the Lindblad equation, the latter is formulated based on a Markovian dynamics that can incorporate the loss of locality due to pure dephasing and amplitude damping for each two-level system (either A or B) via the jump operators
\begin{equation}
\hat L_1^{A(B)} = \left( \begin{matrix}
0 & 0 \\
0 & 1
\end{matrix} \right) 
\,, \quad
\hat L_2^{A(B)} = \left( \begin{matrix}
1 & 0 \\
0 & 0
\end{matrix} 
\right)
\end{equation} 

The use of these operators results in dephasing irrespective of whether or not the system is in state $| 0 \rangle$ or state $| 1 \rangle$ of the Hamiltonian basis or in some superposition of these states. In other words, each operator contributes to the decay of the off-diagonal elements of the density matrix without energy exchange for the case of $\hat L_1^{A(B)}$. This choice of operators is employed in the Lindblad equation, which is expressed as
\begin{equation}
\frac{d\hat \rho}{dt} = -\frac{i}{\hbar} \left[ \hat H , \hat \rho \right] + \gamma \mathcal{L} ( \hat \rho )
\end{equation}
where the coefficient $\gamma$ represents the strength of the interaction and the operator $\mathcal{L} (\hat \rho)$ depends on the jump operators as follows:
\begin{equation}
\mathcal{L} (\hat \rho) = \sum_{m=1}^2 \hat L_m \hat \rho \hat L_m^{\dagger} - \frac{1}{2} \{ \hat L_m^\dagger \hat L_m , \hat \rho \} \,.
\end{equation}

To incorporate the dynamics of the two subsystems \textit{A} and \textit{B}, the general density matrix is given by $\hat \rho = \hat{\rho}_A \otimes \hat{\rho}_B$ and initially assumed to be a Bell diagonal state that once perturbed evolves in time without generating new correlations. The density operators for each subsystem are then found from partial traces of $\hat{\rho}$ and the jump operators for each subsystem constructed. The global jump operator is given by the Kronecker product $\hat L_i = \hat L_i^{A} \otimes \hat L_i^{B}$.

\subsection{Equilibrium and Stationary States}

In the SEAQT framework, any state whose entropy does not evolve in time (i.e., $dS/dt = 0$) is a non-dissipative state. In Eq.~(\ref{EOM9}), this corresponds to $D\left(\hat{\rho}\right)/Dt = 0$ provided there are no reversible dynamics, i.e., $\hat{\rho}$ commutes with the Hamiltonian $\hat{H}$. This is the case for every stable equilibrium state. When the only generators of the motion are the identity and Hamiltonian operators, the density operator for such states is defined by the canonical distribution $\hat{\rho}_{eq} = \textrm{exp}(-\beta\hat{H})/Z$.  

Other states in which there is no dissipation are so-called partially canonical states of a simple quantum system \cite{beretta_nonlinear_2006}, which are not stable equilibrium states, i.e., they are not globally stable but instead only locally so. This is the case for unstable or metastable equilibrium states \cite{gyftopoulos_thermodynamics_2012} or for maximally entangled states such as Bell diagonal states. Composite quantum systems may experience partially canonical states where the reversible dynamics of the system are still present but the dissipation is zero. As an example, consider the spectral representation of the composite system density operator given by
\begin{equation}
\hat{\rho} = \sum_i\lambda_i \hat{P}_i
\end{equation}
where the $\lambda_i$ are the density operator eigenvalues and the $\hat{P}_i$ are the associated projectors constructed from the Bell states
\begin{align}
    | \Phi^\pm \rangle &= \frac{1}{\sqrt{2}} \left( 
| 0 0 \rangle \pm | 1 1 \rangle
    \right) \\ \nonumber
  | \Psi^\pm \rangle &= \frac{1}{\sqrt{2}} \left( 
| 0 1 \rangle \pm | 1 0 \rangle
    \right) 
\end{align}

Now, if the system is assumed to be bipartite where both subsystems $A$ and $B$ have two energy eigenlevels and the composite system state is a Bell diagonal state, the four projectors can be written as
\begin{equation}
\hat{P}_{\Phi^\pm} = |  \Phi^\pm \rangle \langle {\Phi^\pm} | =
\frac{1}{2} 
\begin{bmatrix}
1 & 0 & 0 & \pm 1 \\
0 & 0 & 0 & 0 \\
0 & 0 & 0 & 0 \\
\pm 1 & 0 & 0 &  1 \\
\end{bmatrix}
\end{equation} 
\begin{equation}
\hat{P}_{\Psi^\pm} = | \Psi^\pm \rangle \langle \Psi^\pm | = 
\frac{1}{2}
\begin{bmatrix}
0 & 0 & 0 & 0 \\
0 & 1 & \pm 1 & 0 \\
0 & \pm 1 & 1 & 0 \\
0 & 0 & 0 & 0 \\
\end{bmatrix}
\end{equation} 

To obtain a representation of the state for each subsystem, the partial trace of the composite can be found such that
\begin{equation}
\hat{\rho}_J = \Tr_{\bar{J}}\left(\hat{\rho}\right) = \sum_i\lambda_i \Tr_{\bar{J}}\left(\hat{P}_i\right) = \sum_i\lambda_i \hat{I}_J/2 = \hat{I}_J/2
\end{equation}
where $J = A,B$ and $\bar{J} = B,A$. Thus, for any Bell diagonal state, the resulting density operator for subsystem $A$ or $B$ is maximally mixed. Noting that any function of a density operator can be written as a function of its eigenvalues, i.e.,
\begin{equation}
f\left(\hat{\rho}\right) =  \sum_i f\left(\lambda_i\right) \hat{P}_i
\end{equation}
it follows that for any Bell diagonal state, the partial trace of any function of the density operator (e.g., $\hat{B}\ln(\hat{\rho})$ where $\hat{B}$ is the idempotent operator) will be proportional to the identity operator. Next, theorem 7 of \cite{beretta_quantum_1985} states that any density operator satisfying the following relationship for all $J$ is a non-dissipative state of the composite system for which the generators of the motion are $\hat{I}$ and $\hat{H}$:
\begin{equation}
\hat{\rho}_J\left(\hat{B}\ln(\hat{\rho})\right)^J = \hat{\rho}_J\left[\lambda_{IJ}\hat{I}_J + \lambda_{HJ}\hat{H}\right]
\label{Eq:Theorem_7}
\end{equation}
where $\lambda_{IJ}$ and $\lambda_{HJ}$ must be real scalars. Evaluating $\left(\hat{B}\ln(\hat{\rho})\right)^J$ for a Bell diagonal state results in
\begin{eqnarray}
\left(\hat{B}\ln(\hat{\rho})\right)^J = \Tr_{\bar{J}}\left(\left(\hat{I}_J\otimes\hat{\rho}_{\bar{J}}\right)\hat{B}\ln(\hat{\rho})\right) \nonumber \\ = \Tr_{\bar{J}}\left(\hat{B}\ln(\hat{\rho})\right) = \alpha\hat{I}_J
\end{eqnarray}
where $\alpha$ is a real scalar dependent upon the eigenvalues of the Bell diagonal state. Thus, all Bell diagonal states satisfying Eq. (\ref{Eq:Theorem_7}) are non-dissipative, yet they are not stable equilibrium states. Therefore, to analyze the change of entanglement of states related to Bell diagonal states using the SEAQT equation of motion, it is necessary to use a perturbation method to find dissipative states that have features similar to Bell diagonal states. \\

In the case of the Lindblad framework, its equation of motion does not evolve to stable equilibrium (i.e., because of the choice of jump operators used here, there is no information in this equation to guarantee compatibility with the unique canonical states of equilibrium thermodynamics) but instead to some stationary state, which may, as it is in the present case, be independent of the initial state. This, of course, also depends on the choice of jump operators made.  Thus, the stationary state is unique provided 
\begin{equation}
\mathcal{L}( \hat \rho_{SS})  = 0
\end{equation}
which implies that the semi-group relaxes as long as the zero eigenvalues of $\mathcal{L}$ are non-degenerate and the remaining ones have a negative real part. If not, the final stationary state may depend on the initial state \cite{de_vega_dynamics_2017}. In either case, within the framework of thermodynamics, this stationary state is dissipative and not canonical.

\subsection{Perturbation Approaches}
\label{Sec:Perturb}

The initial density operator $\hat{\rho}_0$ that is perturbed here is the one for the Bell diagonal state in the c-configuration space given by
\begin{equation}
\hat{\rho}_0 = \frac{1}{4}\left(\hat{I}_A\otimes\hat{I}_B + \sum_{i = 1}^3 c_i \hat{\sigma}_{iA}\otimes\hat{\sigma}_{iB}\right)
\end{equation}
where the $\hat{\sigma}_{iA(B)}$ are the Pauli \textit{x}, \textit{y}, and \textit{z} matrices, respectively, $|c_i| \leq 1$ are real scalar coefficients. In order to reproduce the experimental states in \cite{liu_time-invariant_2016}, the scalar coefficient values used here are $c_1 = 0.996$, $c_2 = 0.4$, and $c_3 = -0.4$ although $c_1$ is also varied parametrically in some of the figures. 

This section presents two distinct approaches for generating perturbed states from $\hat{\rho}_0$ where each perturbation approach produces perturbed states that share some particular feature with $\hat{\rho}_0$. 

\subsubsection{Pure State Weighted-Average Approach}
The first proposed approach is a simple weighted average of $\hat{\rho}_0$ with the density operator, $\hat{\rho}_{\text{pure}}$, of a pure state of zero energy whose subsystems both also have zero energy. In a sense, this perturbation approach is compatible with a constrained perturbation with only energy as the constraint. The following expression is used to generate a perturbed state $\hat{\rho}_0'$ with this method: 
\begin{equation}
\hat{\rho}_0' = \zeta \hat{\rho}_0 + \left(1 - \zeta\right)\hat{\rho}_{\text{pure}}
\end{equation}
where
\begin{equation}
\hat{\rho}_{\text{pure}} \equiv \hat{\rho}_{A,\text{pure},x}\otimes\hat{\rho}_{B,\text{pure},x} = \frac{1}{4}
\begin{bmatrix}
1 & 1 \\ 1& 1
\end{bmatrix}
\otimes
\begin{bmatrix}
1 & 1 \\ 1& 1
\end{bmatrix}
\end{equation}
Here the subscript $x$ denotes that the polarization vector of $\hat{\rho}_{A(B),\text{pure},x}$ points along the \textit{x}-axis of the Bloch sphere. The weighting parameter $\zeta$ ranges from 0 to 1. This method is used because the resulting perturbed state density operator, $\hat{\rho}_0'$, has the same energy as $\hat{\rho}_0$ (though only one particular value of $\zeta$ ensures that $\hat{\rho}_0'$ has the same entropy as $\hat{\rho}_0$) and also because no new entanglement is introduced into the perturbed state. Thus, while $\hat{\rho}_0'$ is not a Bell diagonal state, this guarantees that the stable equilibrium state to which $\hat{\rho}_0'$ evolves is a Bell diagonal state although one different than the original Bell diagonal state.

\subsubsection{General Bipartite System Approach}

The second perturbation approach is the general bipartite perturbation scheme \cite{montanez-barrera_method_2022,holladay_steepest-entropy-ascent_2019}. This perturbation method enables the perturbation of an arbitrary density operator, while preserving its unit trace, energy, entropy, and other physical constraints. The algorithm begins by calculating the non-negative square root of the initial density operator as
\begin{equation}
\hat{\gamma}_0 = \sqrt{\hat{\rho}_0},
\end{equation}
after which the perturbed square root is constructed by perturbing $\hat{\gamma}_0$ according to
\begin{equation}
\hat{\gamma}_\epsilon = \hat{\gamma}_0 + \frac{1}{2}\sum_{i,j=0}^3 \eta_{ij} \hat{\sigma}_i^A \otimes \hat{\sigma}_j^B,
\end{equation}
where $\hat{\sigma}_i$ represents the set  of Pauli matrices $ \hat{I}$, $\hat{\sigma}_x$, $\hat{\sigma}_y$ and $\hat{\sigma}_z$, the superscripts $A$ and $B$ denote the subsystems to which the operators pertain, and $\eta \in$ GUE with mean zero and standard deviation $\sigma$
\begin{equation} \label{eq:seteq}
\Tr \left( \hat{\gamma}_r^2 \hat{C}_i (\hat{\gamma}_r) \right) = \Tr \left( \hat{\rho}_0 \hat{C}_i (\hat{\gamma}_0) \right),
\end{equation}
where the $C_i$ are the ${ \hat{I}, \hat{H}, -\hat{B} (\hat{\gamma}^2 - \hat{I}) }$ operators. Eq. (\ref{eq:seteq}) thus, maintains constant unit trace, energy, and linear entropy respectively. Furthermore,
\begin{equation}
\hat{\gamma}_r = \hat{\gamma}_\epsilon - \sum_{i=1}^3 \{ \hat{G}_i (\hat{\gamma}_\epsilon), \hat{\gamma}_\epsilon \} \lambda_i,
\end{equation}
where the $\lambda_i$ coefficients are derived from the constraint equations, Eq. (\ref{eq:seteq}), and the operators $\hat{G}_i$ are part of the symmetrized gradients $\{ \hat G_i (\hat \gamma^2) , \hat \gamma \}$ with $\hat G_1 = \hat I$, $\hat G_2 = \hat H$ and $\hat G_3 (\hat \gamma )= -(2 \hat \gamma^2 -I)$. The resulting set of equations contains 12 roots according to the classical Bezout theorem, and only the solution with $E\left( \hat \rho \right)$ close to the Bell diagonal state is kept, i.e., that perturbed state with entanglement and separability close to the original Bell diagonal state.

\section{Results}
\label{Sec:Results}

This section presents results for the evolution of system entanglement using the SEAQT and Lindblad equations of motion starting from perturbed states. In addition, various correlations between the properties of the perturbed and stable equilibrium states are presented for the SEAQT framework. First, Section \ref{Sec:weighted} shows entanglement evolutions using the weighted average perturbation approach to generate the initial states used by the SEAQT equation of motion as well as the Lindblad equation of motion. 
Next, Section \ref{Sec:constrained} does likewise except that the initial states used are generated by the general bipartite perturbation approach with unit trace and constant energy and entropy constraints and a perturbation matrix with a standard deviation of $\sigma = 10^{-1}$. This arbitrary value is chosen to align with experimental values, but alternative selections could be employed.\\
\begin{figure} [!htb]
	\centering
		a) \includegraphics[scale=\FigScaleFact]{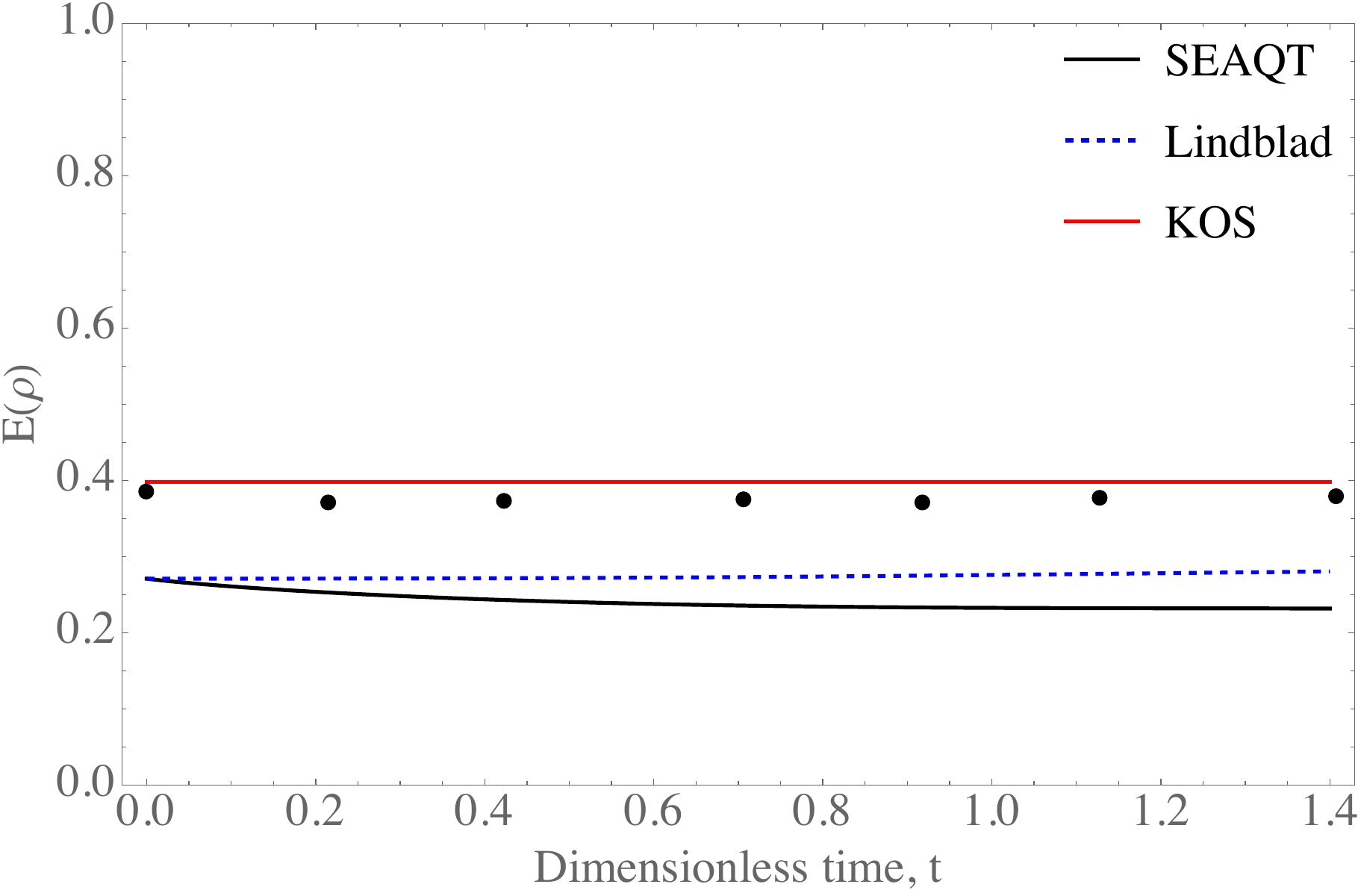} \\
		b) \includegraphics[scale=\FigScaleFact]{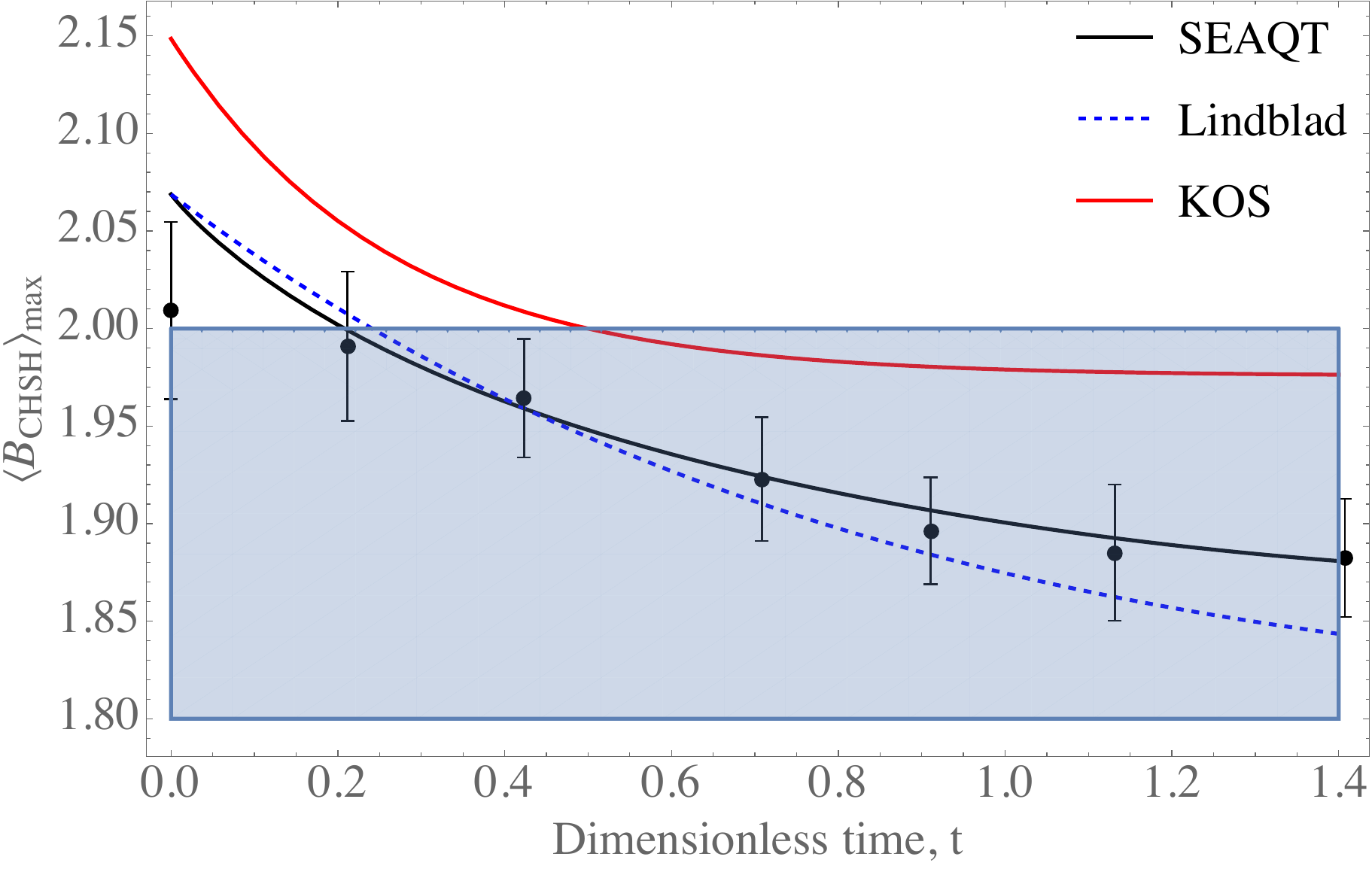}
	\caption{Theoretical and experimental a) concurrence and b) $\langle{\hat{B}_{CHSH}\rangle}_{\mx}$ evolutions for the Bell diagonal state characterized by the parameters $c_1 = 1$, $c_2 = 0.4$, and $c_3 = -0.4$; the experimental (black dots) and KOS theoretical (red curve) results of \cite{liu_time-invariant_2016} were generated using a quantum optics experiment and the Kraus-Operator-Sum approach; the initial state for the SEAQT and Lindblad evolutions is generated using the weighted-average perturbation method with $\zeta=0.68$.}
	\label{Fig::AllBC_SEAQT}
\end{figure}
To gain a qualitative understanding of the effects of dissipation due to an external environment on the entanglement properties of the system, theoretical and experimental results developed by Liu $et\; al.$ \cite{liu_time-invariant_2016} for the evolution of a Bell diagonal state in a stochastic global dephasing (or decoherence) field are shown in Fig. \ref{Fig::AllBC_SEAQT}. Their theoretical results generated using the Kraus-Operator-Sum approach predict that the concurrence remains constant, while $\langle{\hat{B}_{CHSH}\rangle}_{\mx}$ decays. Thus, after a finite time (dimensionless time of $t \approx 0.58$) the theoretical results suggest that the system undergoes so-called sudden death of non-locality, i.e., $\langle{\hat{B}_{CHSH}\rangle}_{\mx}$ goes below 2. The experimental results produced by Liu $et\; al.$ qualitatively validate the theoretical results since the measured concurrence remains roughly constant, while the measured $\langle \hat{B}_{CHSH}\rangle_{\mx}$ decays through time although sudden death of non-locality occurs at a dimensionless time of $t \approx 0.2$. 

Fig. \ref{Fig::AllBC_SEAQT} also presents results for the evolution of $E(\hat \rho )$ and $\langle \hat B_{CHSH} \rangle_{\mx}$ using the SEAQT and Lindblad equations of motion and a weighting parameter of $\zeta = 0.68$ to generate the initial state using the first perturbation method. As can be seen, the sudden death of non-locality predicted with the SEAQT as well as the Lindblad equation matches the experimental result better than the model proposed by Liu $et\; al.$, predicting this death of non-locality at times near 0.2. However, using the weighted-average perturbation method, the $E(\hat \rho )$ is significantly below the 0.4 reported by the experimental results. The discrepancy is due to the weighting parameter which moves the perturbed state away from the Bell diagonal state by nearly 30\% in the direction of a pure state. In Section \ref{Sec:weighted} the evolution of this entanglement measure and the other and their relation to the entropy as well as the entropy generation are explored more systematically, indicating a correlation between thermodynamics and both entanglement measures. 

\subsection{Entanglement Evolution Initialized Via a Weighted-Average Perturbation}
\label{Sec:weighted}
\subsubsection{SEAQT framework}
\label{Sec:SEAQT_weighted}
This section presents evolutions for i) the system concurrence, $E(\hat{\rho})$, and the maximum expectation value of the CHSH operator, $\langle{\hat{B}_{CHSH}\rangle}_{\mx}$; ii) the dependence of the relative entropy of the perturbed initial state and the final stable equilibrium state on the weighting parameter $\zeta$ and the coefficient $c_1$; iii) the dependence of the perturbed state and the final stable equilibrium state entropy on the coefficient $c_1$; iv) the dependence of the stable equilibrium state entropy on the weighting parameter $\zeta$ and the coefficient $c_1$; and v) the dependence of the total dissipation (entropy generation) on the weighting parameter $\zeta$ and the coefficient $c_1$.

\begin{figure}  [!htb]
	\centering
		a)\includegraphics[scale=\FigScaleFact]{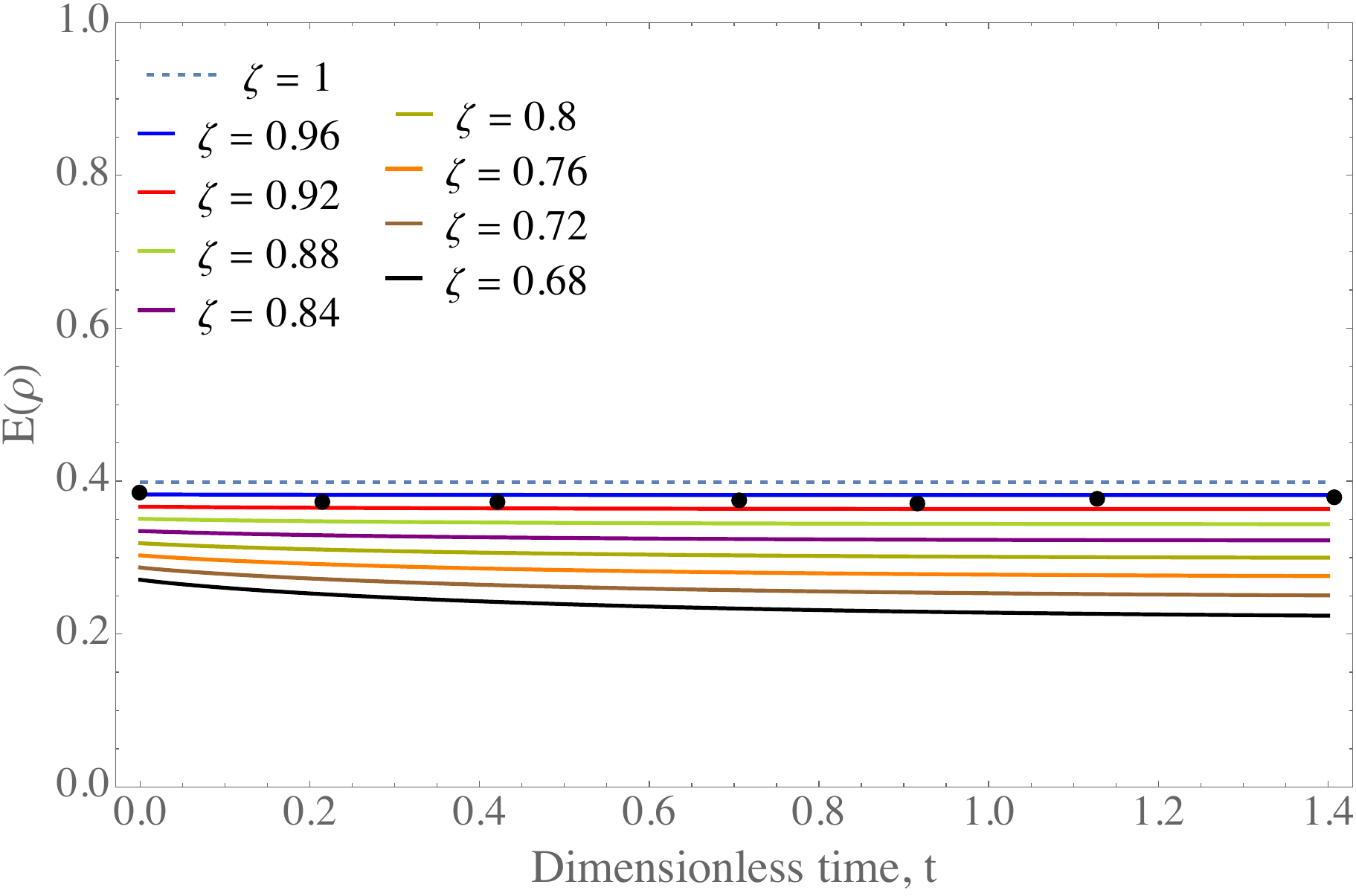} \\
		b) \includegraphics[scale=\FigScaleFact]{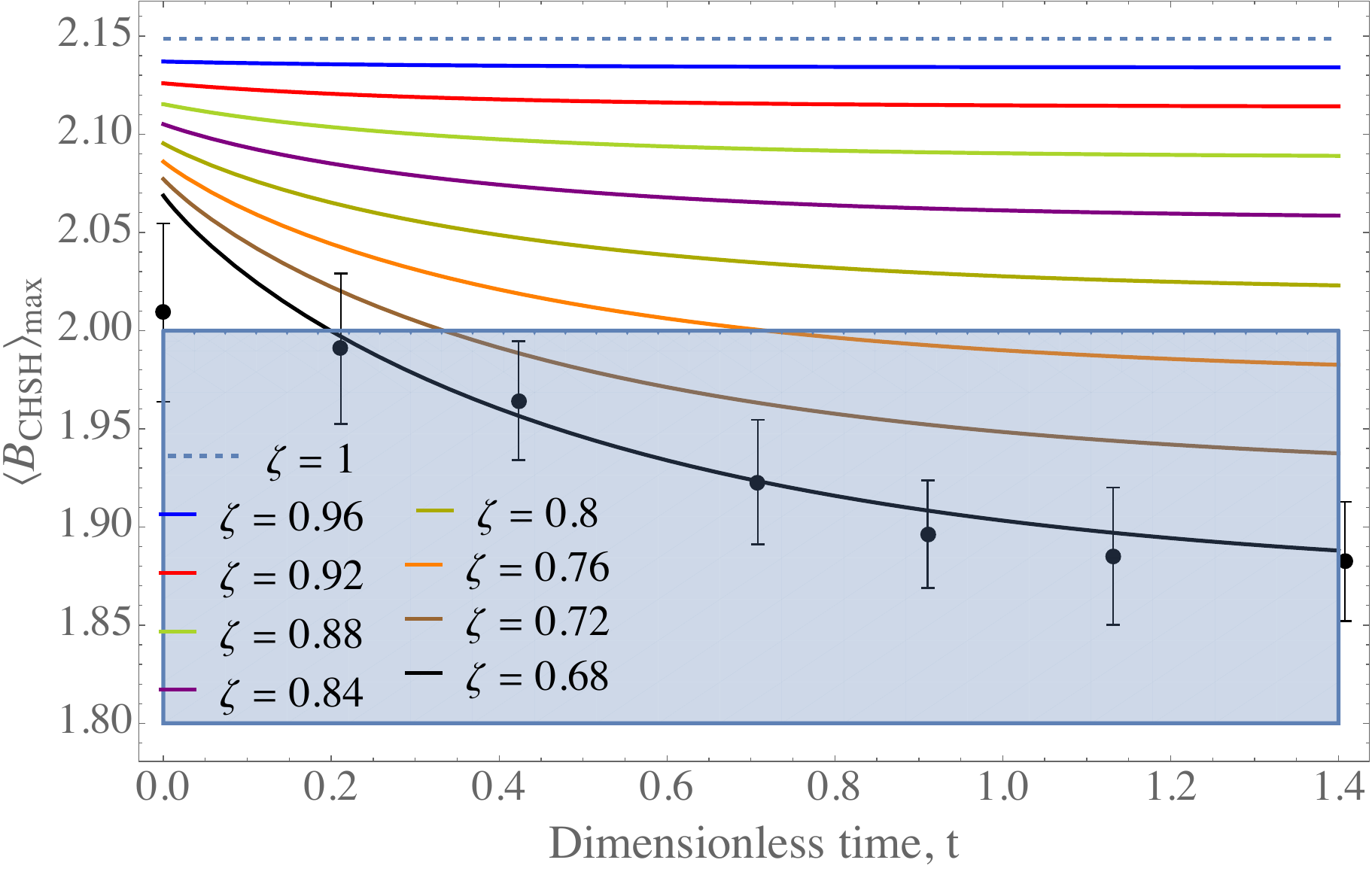}
	\caption{a) $E(\hat{\rho})$ and b) $\langle{\hat{B}_{CHSH}\rangle}_{\mx}$ evolutions in the SEAQT framework for values of the weighting parameter $\zeta = 1$, $0.96$, $0.92$, $0.88$, $0.84$, $0.8$, $0.76$, $0.72$, and $0.68$ where $\zeta = 1$ corresponds to the uppermost dashed line and each lower solid line corresponds to the next lower value of $\zeta$; the experimental results (black dots) of \cite{liu_time-invariant_2016} were generated using a quantum optics experiment.}
	\label{Fig:SEAQT_CB}
\end{figure}

Fig. \ref{Fig:SEAQT_CB} shows the evolutions of $E(\hat{\rho})$ and $\langle{\hat{B}_{CHSH}\rangle}_{\mx}$ for nine different values of the weighting parameter $\zeta$. For both properties, initial decay in their value occurs as entropy is generated internal to the system before becoming constant as the system reaches stable equilibrium. In addition to the initial values of these properties dropping as $\zeta$ decreases, the overall decay in the value of each property increases as $\zeta$ decreases. Fig. \ref{Fig:SEAQT_CB} also shows that for values of $\zeta \approx 0.76$ and below, the composite system undergoes sudden death of non-locality in the way that the Bell diagonal state predicted by Liu $et\; al.$ does. For values above $\zeta \approx 0.76$, the non-locality and, thus, quantum entanglement remain as the system relaxes to stable equilibrium.

Fig. \ref{Fig:SEAQT_CB}, however,  reveals a weakness of using a simple weighted-average perturbation method to generate the initial state used by the SEAQT equation of motion (or for that matter the Lindblad equation (see Section \ref{Sec:Lindblad_weighted} below)). The initial state generated with the smallest weighting factor, $\zeta$, places it closest to a pure (zero-entropy) state and furthest from the entropy of the baseline Bell diagonal state. As $\zeta$ increases, the initial state's entropy approaches that of the Bell diagonal state with the result that the initial $\langle \hat B_{CHSH} \rangle_{\mx}$ value deviates more and more from the initial experimental state value. Conversely, the concurrence moves ever closer to this experimental value. The problem is that even though the entropy is now closer to that of the Bell diagonal state, the energy is not, which is verified in Section \ref{Sec:constrainedSEAQT} below, suggesting that the general bipartite perturbation method be used to generate the initial state, since it constrains both the entropy and the energy to be that of the initial Bell diagonal state. 

\begin{figure}  [!htb]
		a) \includegraphics[scale=\FigScaleFact]{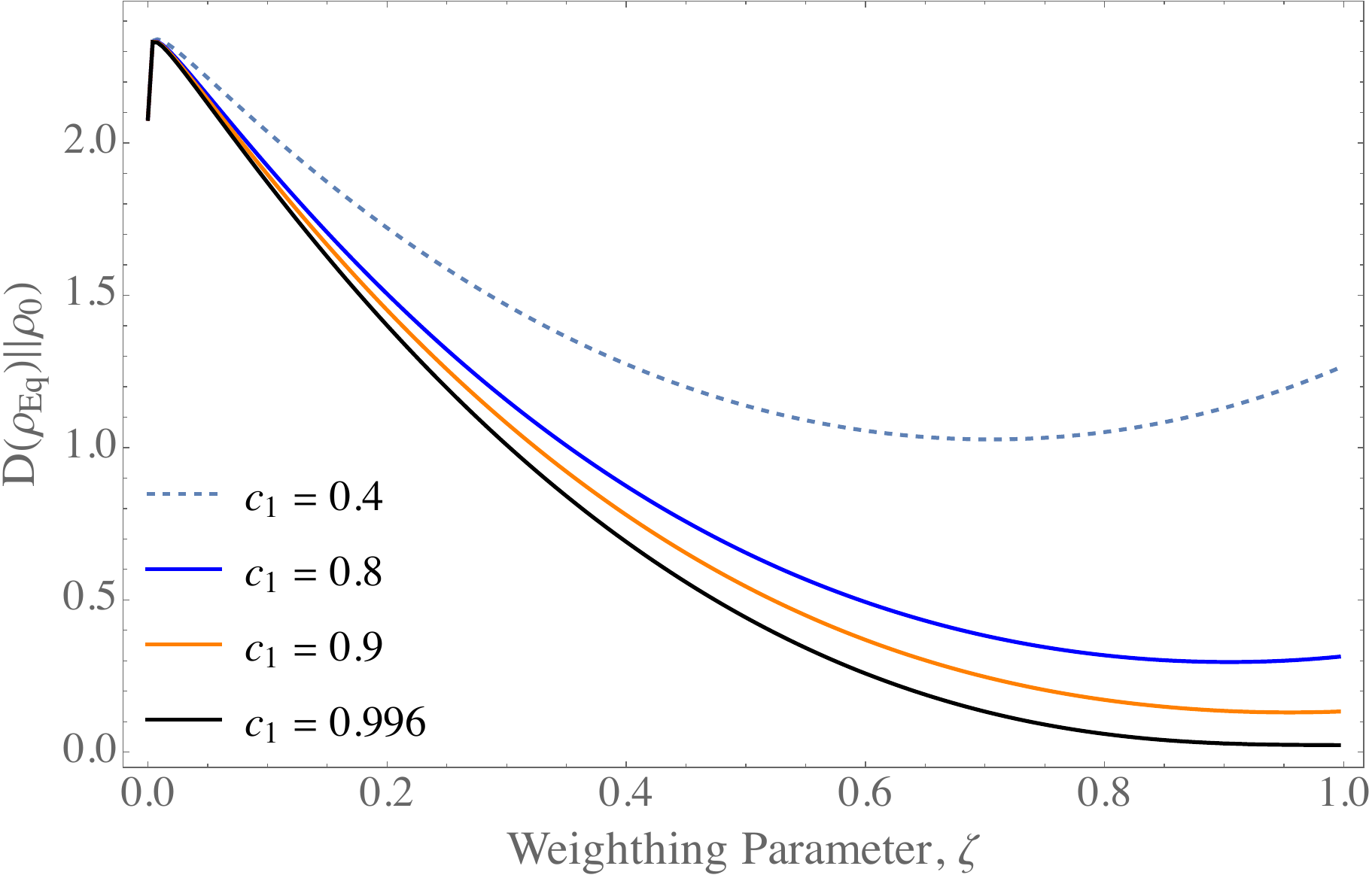} \\
		b) \includegraphics[scale=\FigScaleFact]{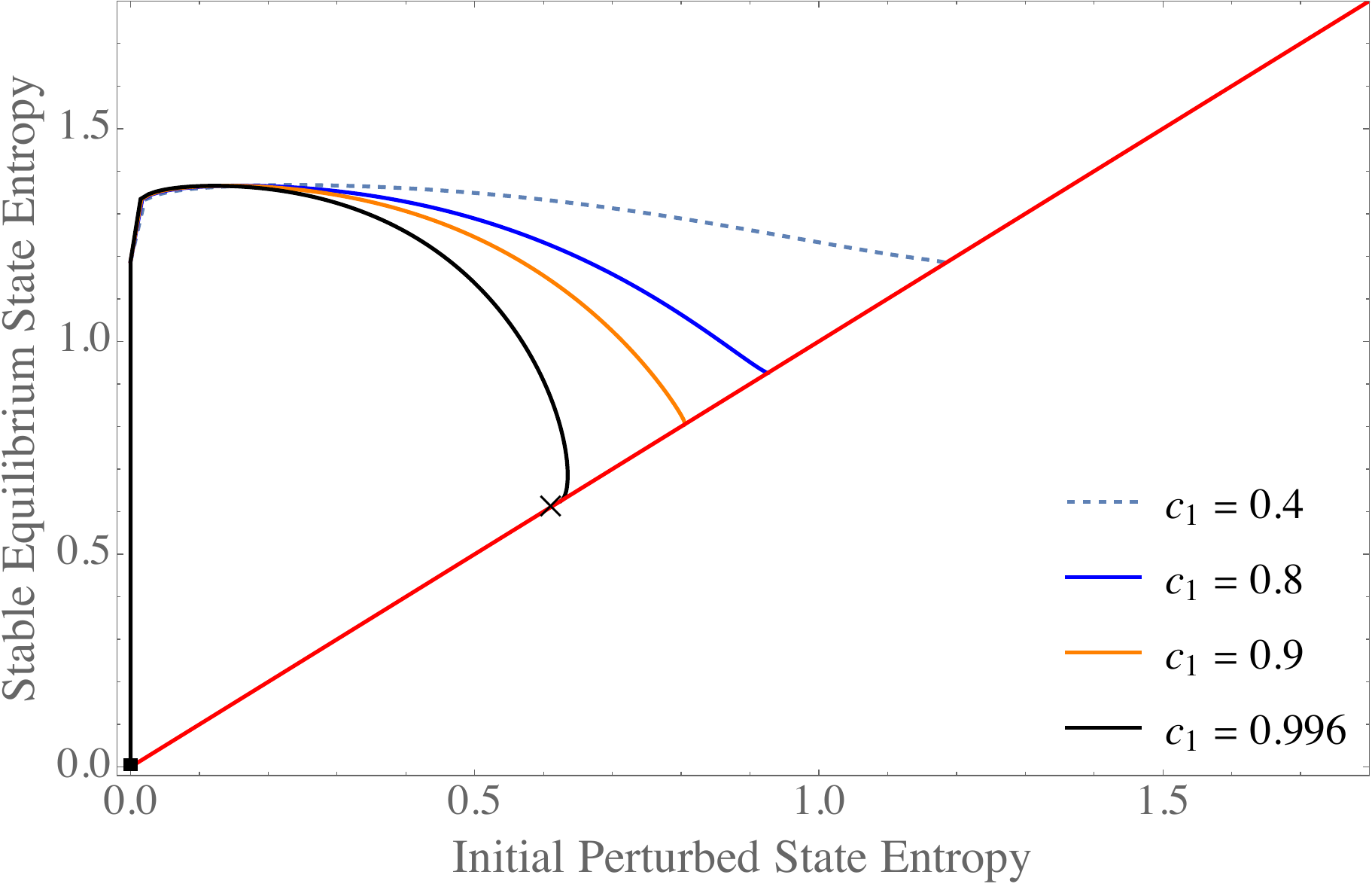}
	\caption{Plots of a) the relative entropy in the SEAQT framework between the initial perturbed Bell diagonal state (for $c_1 = 0.4, \; 0.8, \; 0.9, \; \text{and} \; 0.996$) and the final stable equilibrium state entropy as a function of the perturbation parameter $\zeta$ and b) the initial perturbed Bell diagonal state entropy (for $c_1 =  0.4, \; 0.8, \; 0.9, \; \text{and} \; 0.996$) versus the final stable equilibrium state entropy. The red line is simply a line drawn through the entropies for the unperturbed Bell diagonal states for $c_1 = 0.4, \; 0.8, \; 0.9, \; \text{and} \; 0.996$ whereas the $\blacksquare$ marker indicates the pure state entropy and the $\times$ marker indicates the Bell diagonal state entropy for $c_1=1.0$.}
	\label{Fig:Re_ent} 
\end{figure}

\begin{figure} 	 [!htb]
		\includegraphics[scale=\FigScaleFact]{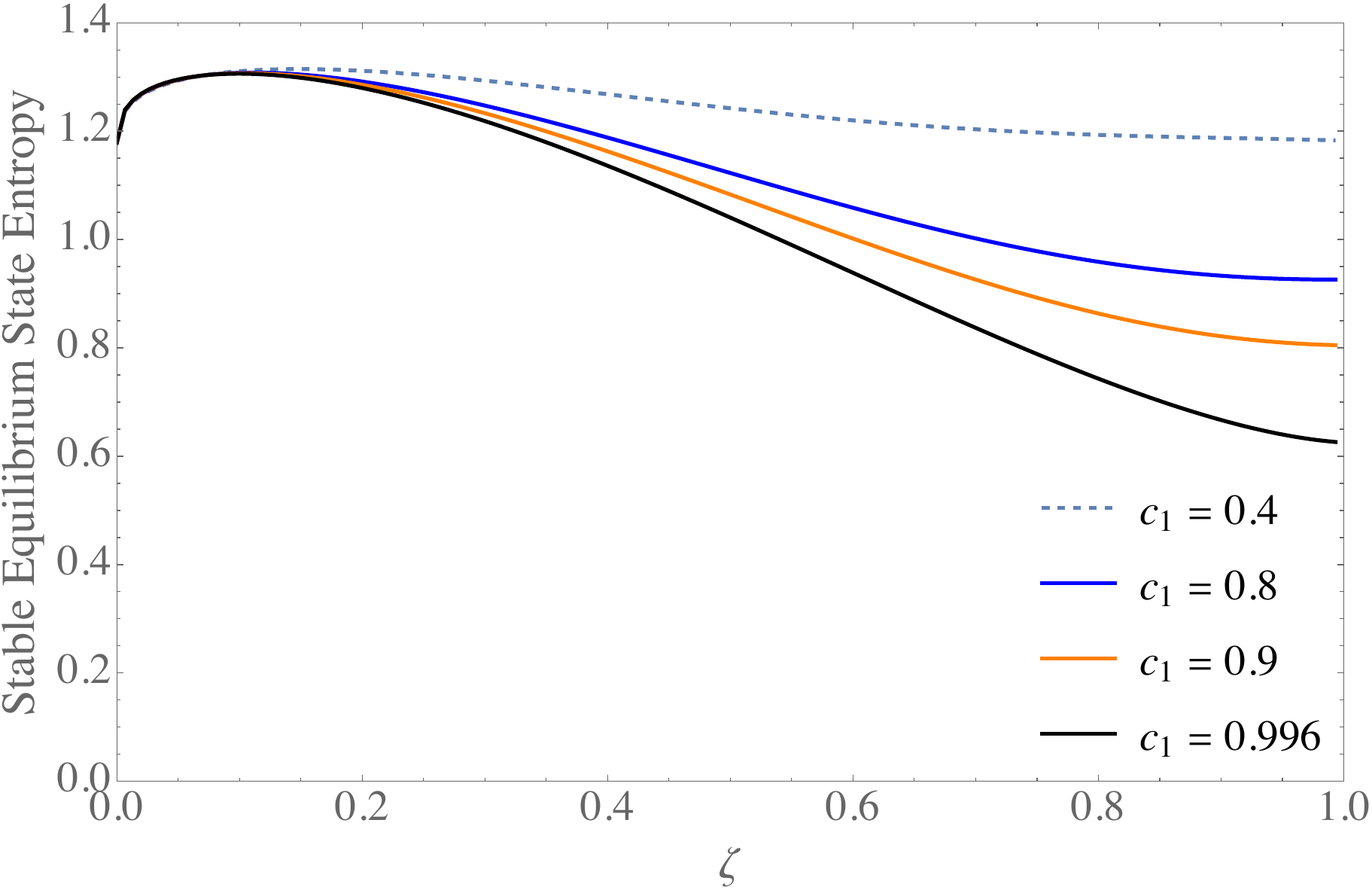} 
	\caption{Dependence in the SEAQT framework between the weighting parameter $\zeta$ and the final stable equilibrium state entropy  for $c_1 = 0.1, 0.4, 0.8, 0.9$ and $0.996$.}
	\label{Fig:SS_EG}
\end{figure}

Fig.~\ref{Fig:Re_ent} a) shows the change of the relative entropy between the initial perturbed state and the final stable equilibrium state (i.e., the so-called Gibbs state) predicted by the SEAQT equation of motion versus the weighting parameter $\zeta$, which varies between zero and one in increments of 0.01 and for different values of the coefficient $c_1$. The Bell diagonal state corresponds to a
$\zeta = 1$ whereas a $\zeta = 0$ corresponds to the pure state. This final state is consistent with the second law of thermodynamics, while as will be seen below that predicted by the Lindblad equation of motion is not. As to the trends seen in Fig.~\ref{Fig:Re_ent} a), the relative entropy initially increases at very small values of $\zeta$ when the initial perturbed state is furthest from the Bell diagonal state and closest to that of the pure state after which it monotonically decreases and then either plateaus or increases. In addition, the different initial Bell diagonal states defined here by changes in the scalar coefficient $c_1$ result in smaller decreases in the relative entropy as the value of $c_1$ decreases. 

Fig.~\ref{Fig:Re_ent}b) provides the final stable equilibrium state entropy of each evolution versus the initial perturbed state entropy for different choices of the $c_1$ parameter. The right part of the curves shown in this figure represent states with $\zeta$ closer to 1 whereas the left part corresponds to states with $\zeta$ closer to 0. In addition, the red line shown is simply a line drawn through the entropies for the unperturbed Bell diagonal states for $c_1 = 0.4, \; 0.8, \; 0.9, \; \text{and} \; 0.996$ whereas the $\blacksquare$ marker indicates the pure state entropy and the $\times$ marker indicates the Bell diagonal state entropy for $c_1=1.0$. Note once more that the Bell diagonal states are non-dissipative in the SEAQT framework as shown via Eq. (\ref{Eq:Theorem_7}). As to the behavior seen in Fig.~\ref{Fig:Re_ent}b), the total dissipation, which occurs as the system evolves to its final stable equilibrium state, monotonically increases to some maximum the further the initial perturbed state is from the unperturbed Bell diagonal state  after which it monotonically decreases. The maximum occurs closer to the pure state than the Bell diagonal state, suggesting that being in the neighborhood of the former leads to the dissipation decrease. Furthermore, an exception to the initial increase in dissipation occurs for the case when $c_1 = 0.996$ where, in fact,
there is a monotonic decrease since the curve initially curves to the right before curving to the left. 

It is noted here that not only is the final state in the SEAQT framework consistent with the second law of thermodynamics but so is every non-equilibrium state along the path predicted by SEAQT equation of motion since it is driven by the SEA principle at every instant of time. Thus, the dissipation
predicted is that, which would result from the second law. Furthermore, the final stable equilibrium state predicted by the SEAQT equation of motion is consistent with the Lyapunov condition for global stability \cite{beretta_nonlinear_2009} and, thus, non-dissipative. The SEAQT framework also admits of metastable equilibrium states, which are perturbatively unstable (i.e., only locally stable \cite{beretta_nonlinear_2009}). These states are related to the elements of the kernel of the dissipation operator $D (\hat \rho )/Dt$. In general, the distributions for these metastable equilibrium states take the form
 \begin{equation}
 \hat \rho = \frac{\hat B \exp \left( -\beta \hat{H} \right) \hat B}{\text{Tr} \left( \hat B \exp \left( -\beta \hat{H} \right) \right)} \label{Eq:distribution}
 \end{equation}
where  $\hat B = \hat{I} - \hat{P}_{\text{ker}\hat \rho}$ is the idempotent operator or projector onto the range of $\hat \rho$, while $\hat{I}-\hat{B}$ is the projector onto the kernel of $\hat{\rho}$. Note that when $\hat B = \hat{I}$,  Eq. (\ref{Eq:distribution}) reduces to the canonical distribution predicted for the stable equilibrium state by the SEAQT equation of motion. The von Neumann entropy can now be expressed as
 \begin{eqnarray}
 S &=& - \text{Tr} \left( \hat \rho \hat B \ln \hat\rho \right) \\
 &=& - \text{Tr} \left( \hat \rho  \ln ( \hat\rho  + \hat{P}_{\text{ker}\rho} ) \right) \,.
 \end{eqnarray}
It can readily be shown that $S$ is well defined for any $\hat{\rho}$, even one that is singular \cite{beretta_maximum_2010}. This, in particular, is true for the entropy of the limiting case of the pure state reported as a dark square in Fig.~\ref{Fig:Re_ent} b). In other words, for pure states, the null vector, $| \psi \rangle$, can easily be calculated, i.e., $\hat \rho_0 | \psi \rangle = 0$, which leads to the projector $\hat{P}_{\text{ker}\hat \rho} = | \psi \rangle \langle \psi |$ and a value of zero for the von Neumann entropy of the pure state, i.e., 
\begin{align}
S_{\text{Pure}}  &= - \Tr (\hat B \hat \rho_{\text{pure}} \ln (\hat \rho_{\text{pure}}) ) =0
\end{align}
Since the pure state corresponds to a completely known state, the projector projects out the zero eigenvalues leaving only the eigenvalue $\lambda =1$. 

The other limiting case reported in Fig.~\ref{Fig:Re_ent} b) is that of the Bell diagonal state for which the null space depends on the selection of the coefficients $c_i$, and for which all the null vectors are sums of the projectors 
\begin{equation}
\hat P_{\Phi^{\pm}} = | \Phi^{\pm} \rangle \langle \Phi^{\pm} | \,, \hat P_{\Psi^{\pm}} = | \Psi^{\pm} \rangle \langle \Psi^{\pm} |
\end{equation}
Furthermore, the null space can be classified according to its cardinality, which accounts for the number of null vectors for a given selection of the $c_i$. In general, the cardinality of the kernel of $\hat \rho_0$ can be written as the binomial coefficient, i.e.,
\begin{equation}
\text{card} \, \left( \text{ker} \hat \rho \right) = \left( 
\begin{matrix}
\text{dim} \mathcal{H} \\
\text{dim} \left( \text{ker} (\hat \rho) \right)
\end{matrix}
\right)
\end{equation}
The limiting case of  $c_1 =1$ and $c_3 =-c_2$ seen in Fig.~\ref{Fig:Re_ent} b) (the $\times$ marker on the red line of Bell diagonal states) is an element of a set with cardinality 6 whose projector is
$\hat B = \hat{I}- |\Phi^- \rangle \langle \Phi^- | - |\Psi^- \rangle \langle \Psi^- |$ 
and for which the von Neumann entropy is found to be
\begin{align}
S_{\text{eq},0} &= -\text{Tr} ( \hat \rho_0  \hat B \ln \hat{\rho}_0 ) \\
&= \ln 2 -\frac{1}{2}c_2 \text{arctanh}\, c_2  -\ln (1-c_2^2)^{1/2}\,, \\
&= 0.6108 \,.
\end{align}
The state evolutions of the SEAQT equation of motion occur between these two limiting cases, i.e., the non-dissipative Bell diagonal state on the right of Fig. \ref{Fig:Re_ent}b) and the non-dissipative pure state on the left of this figure. Any state in between is dissipative with the exception of the stable or metastable equilibrium states of thermodynamics.


Now, Fig.~\ref{Fig:SS_EG} a) shows the equilibrium state entropy versus the weighting parameter for different values of $c_1$. As can be seen, the larger stable equilibrium entropy is associated with Bell diagonal states with $c_1=0.1$, and as $c_1$ approaches 1, the stable equilibrium state entropy decreases towards the limiting stationary (steady-state) entropy of $S_{\text{eq},0} = 0.6108$. 

\subsubsection{Lindblad framework}
\label{Sec:Lindblad_weighted}
\begin{figure}  [!htb]
	\centering
		a) \includegraphics[scale=\FigScaleFact]{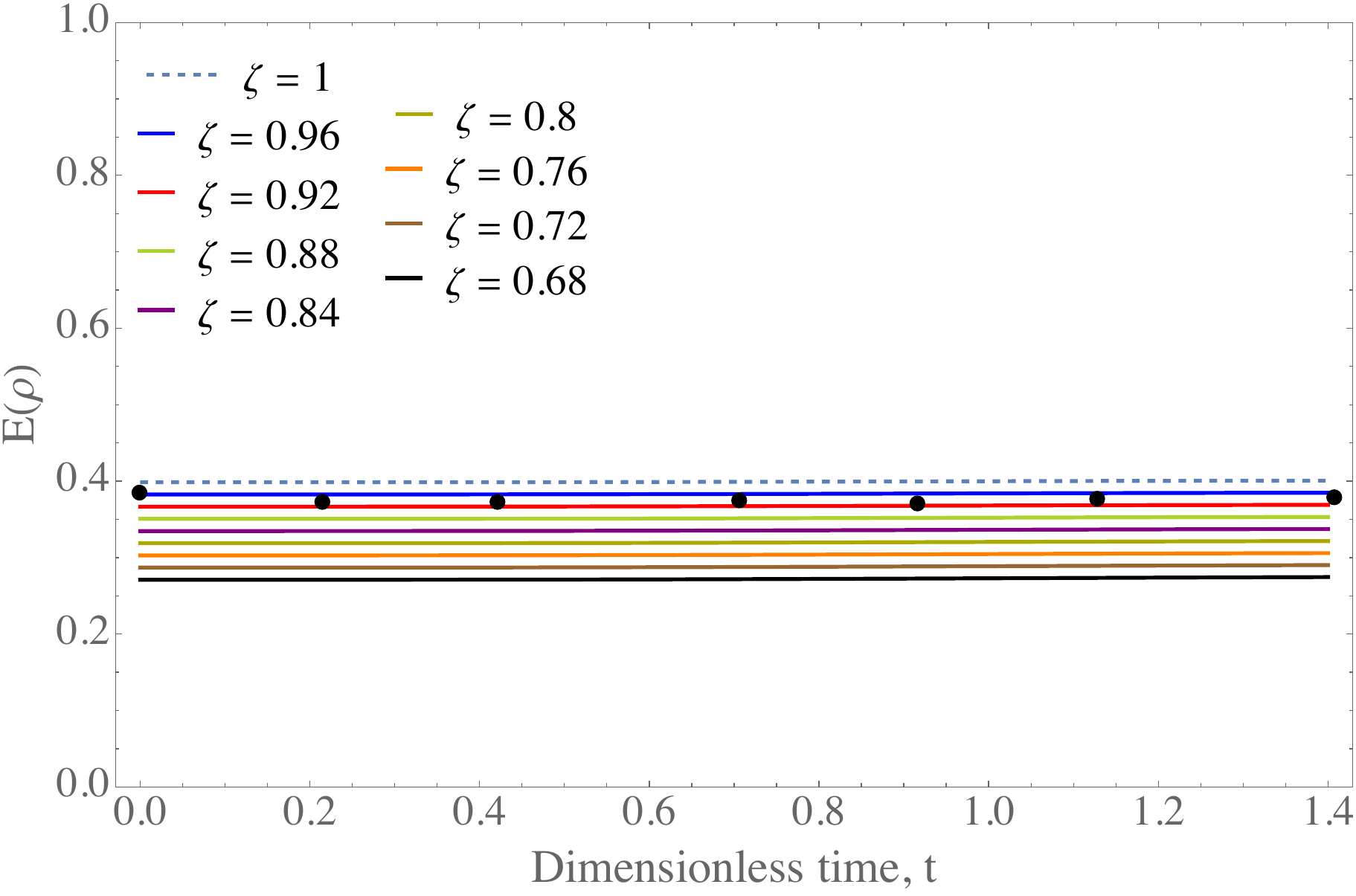} \\
		b) \includegraphics[scale=\FigScaleFact]{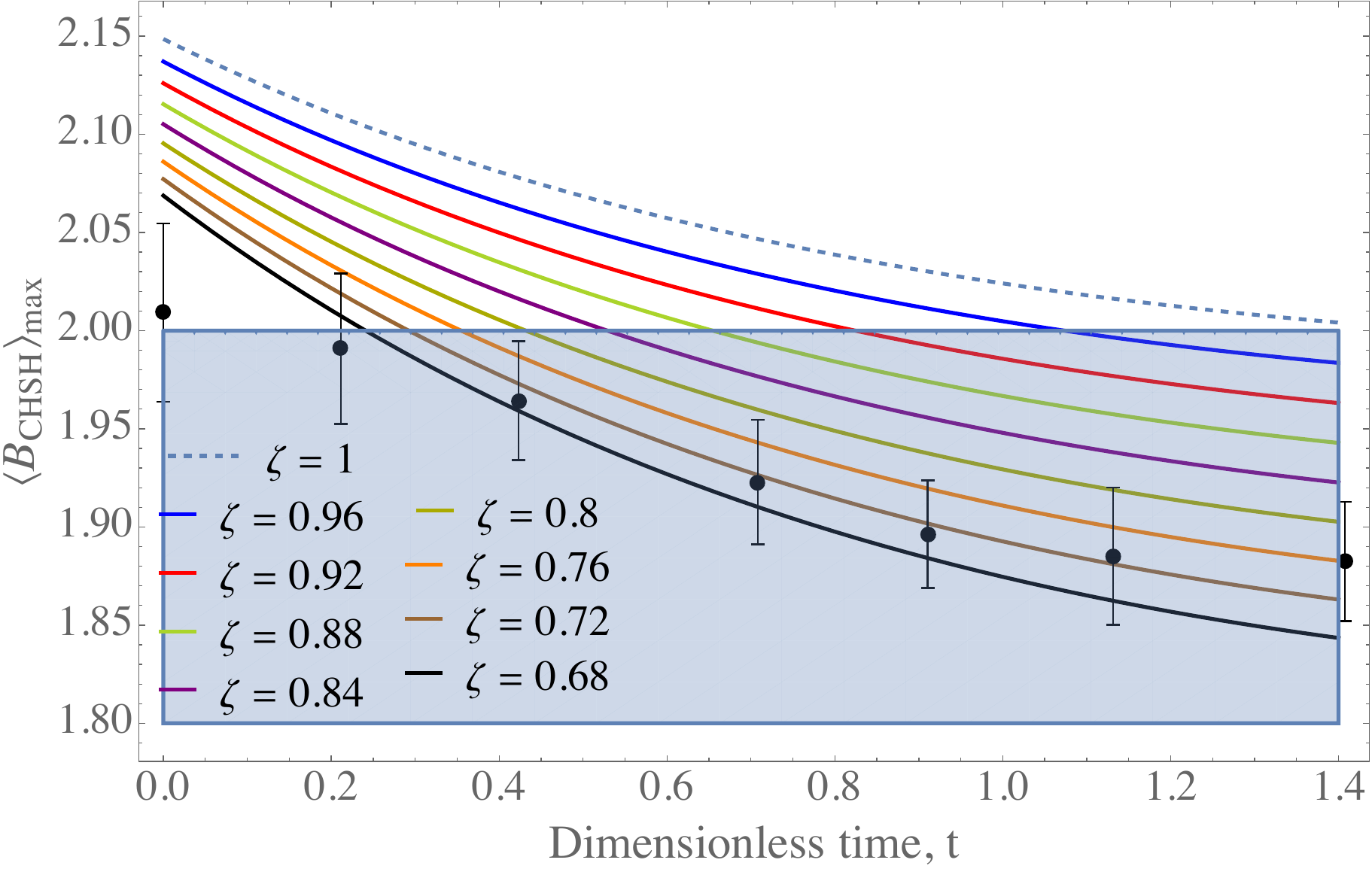}
	\caption{Theoretical and experimental a) concurrence and b) $\langle{\hat{B}_{CHSH}\rangle}_{\mx}$ evolutions in the Lindblad framework for values of the weighting parameter $\zeta = 1$, $0.96$, $0.92$, $0.88$, $0.84$, $0.8$, $0.76$, $0.72$, and $0.68$ where $\zeta = 1$ corresponds to the uppermost dashed line and each lower solid line corresponds to the next lower value of $\zeta$; the experimental results (black dots) of \cite{liu_time-invariant_2016} were generated using a quantum optics experiment.}
	\label{Fig:AllBC_Lind}
\end{figure}

In this section, the dynamics of the Linblad framework are considered. How these dynamics affect the information measures $C(\hat \rho )$ and $\langle \hat B_{CHSH} \rangle_{\mx}$ is examined first. Fig. \ref{Fig:AllBC_Lind} shows nine perturbations employing the weighting parameter. As is seen, in contrast with the SEAQT formulation, the Lindblad dynamics predicts a constant value for the concurrence but like the SEAQT framework a decay of the non-locality measure where the sudden death of non-locality for the $\zeta = 0.68$ curve occurs at $t\approx 0.23$ (as compared to $t\approx 0.19$ for the SEAQT evolution for this value of the weighting parameter).

In general, the entropy changes in the Lindblad framework depend on the Lindblad operator, $\mathcal{L}$, such that
\begin{equation} \label{eq:Ent_Lind}
\frac{d S}{dt} =  \gamma \Tr \, \left( \mathcal{L} \left( \hat \rho \right) \ln \hat \rho \right)
\end{equation}
where the Lindblad operator for the case of Bell diagonal states can be written as
\begin{equation}\label{eq:eq_Lind}
\mathcal{L}\left( \hat \rho_0 \right) =-\frac{1}{4} \gamma ^2\left(
\begin{array}{cccc}
 0 & 0 & 0 &  c_1-c_2 \\
 0 & 0 & 0 & 0 \\
 0 & 0 & 0 & 0 \\
  c_1-c_2 & 0 & 0 & 0 \\
\end{array}
\right)
\end{equation}
A state is not a stationary state of the Lindblad equation except when $c_1 = c_2$. In such cases, the entropy remains constant, and the density matrix's evolution is influenced only by the symplectic term. 

\begin{figure}  [!htb]
		a) \includegraphics[scale=\FigScaleFact]{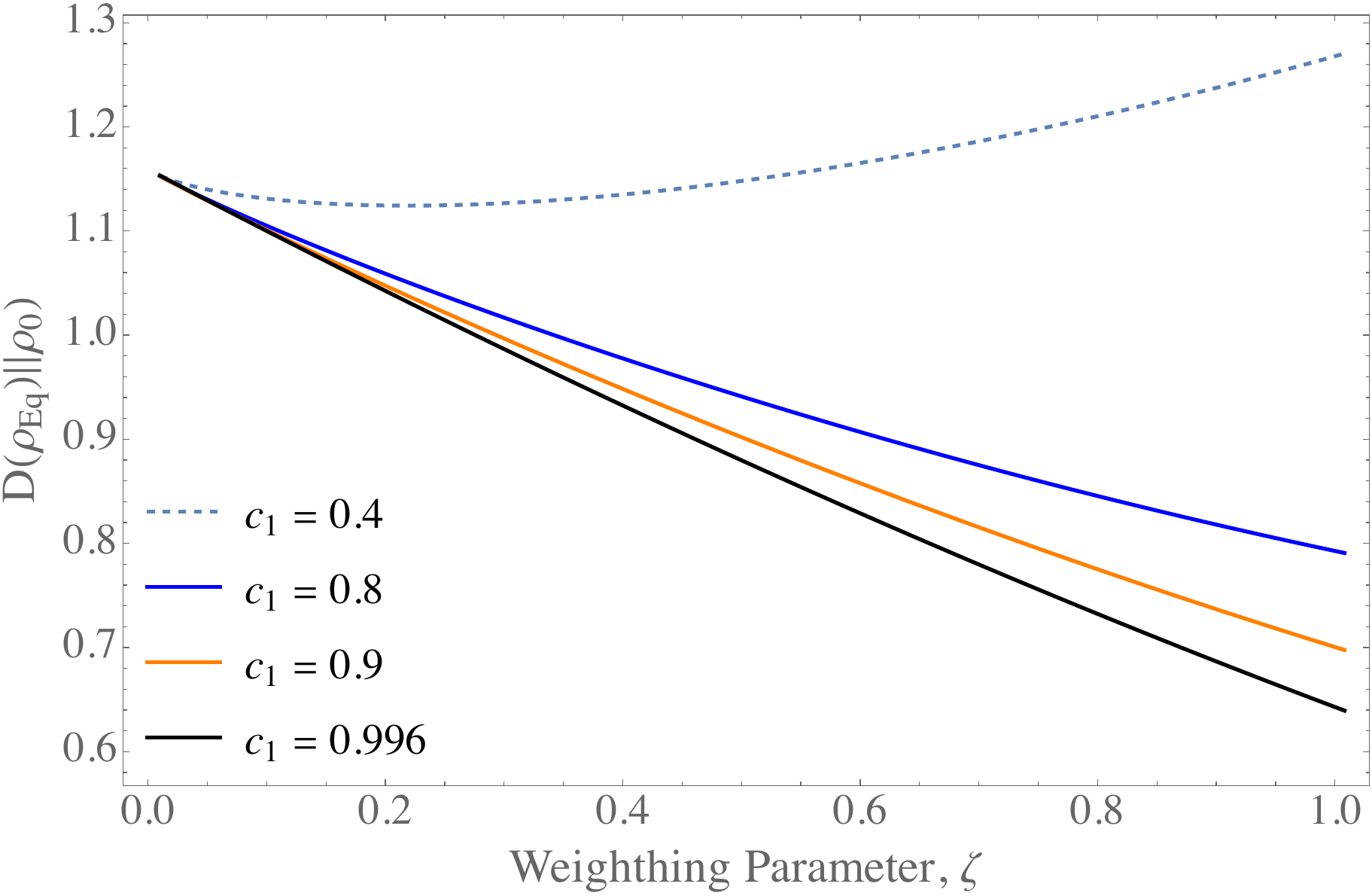} \\
		b) \includegraphics[scale=\FigScaleFact]{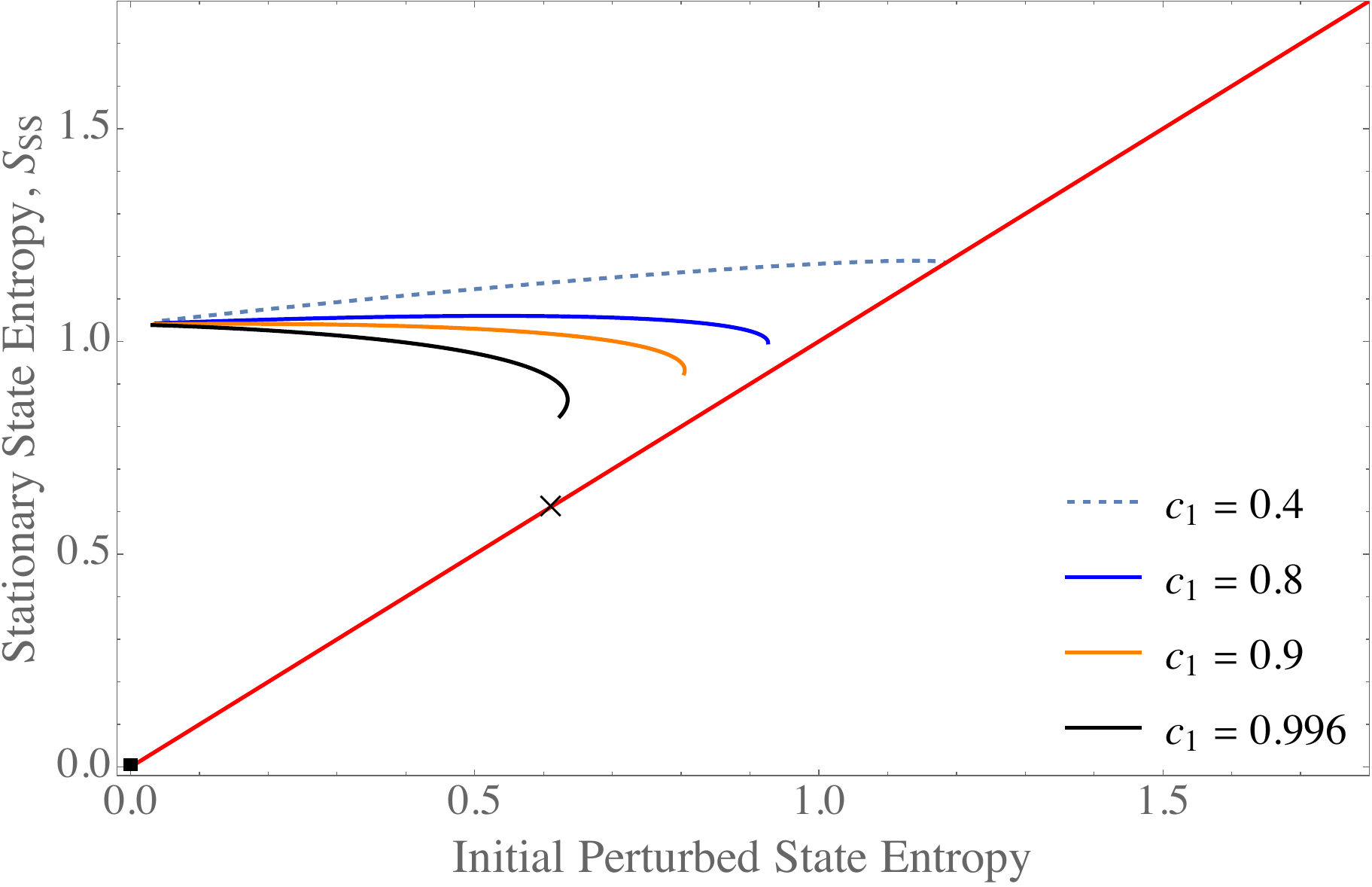}
	\caption{Plots of a) the relative entropy in the Lindblad framework between the initial perturbed Bell diagonal state (for $c_1 = 0.4, \; 0.8, \; 0.9, \; \text{and} \; 0.996$) and the final stationary state as a function of the perturbation parameter $\zeta$ and b) the initial perturbed Bell diagonal state entropy (for $c_1 =  0.4, \; 0.8, \; 0.9, \; \text{and} \; 0.996$) versus the final stationary state entropy. The red line is simply a line drawn through the entropies for the unperturbed Bell diagonal states for $c_1 = 0.4, \; 0.8, \; 0.9, \; \text{and} \; 0.996$ whereas the $\blacksquare$ marker indicates the pure state entropy and the $\times$ marker indicates the Bell diagonal state entropy for $c_1=1.0$.}
	\label{Fig:Re_ent_Lind} 
\end{figure}

Fig.~\ref{Fig:Re_ent_Lind} a) illustrates the change in the relative entropy between the initial perturbed Bell diagonal state entropy and the final stationary state entropy predicted by the Lindblad equation. This final state is not consistent with the second law of thermodynamics. It is instead simply a stationary state that is based on the jump operators chosen to represent the dephasing processes of the experiment modeled here. In general, the Lindblad equation does not recover a final stationary state consistent with the second law unless a quantum channel that leads to a thermodynamic path compatible with the second law obeys the `unital' condition \cite{witten_mini-introduction_2020,reichental_thermalization_2018}. Meeting this condition depends on the Kraus operators used and, as a consequence, on the jump operators chosen. The ones used here do not meet this condition. Thus, this final state is not a stable equilibrium state. As to the trends seen in this figure, the final stationary state entropy approaches that of the initial perturbed Bell diagonal state as $\zeta$ increases with the relative entropy monotonically decreasing for all values of $c_1$ with the exception of $c_1 = 0.4$. For the latter, it decreases to a minimum and then increases. In general, a decrease in $c_1$ leads to a greater deviation of the final stationary state entropy from the initial perturbed state entropy.

Fig.~\ref{Fig:Re_ent_Lind} b) shows the initial perturbed Bell diagonal state entropy (for $c_1 =  0.4, \; 0.8, \; 0.9, \; \text{and} \; 0.996$) versus the final stationary state entropy. The red line is simply a line drawn through the entropies for the unperturbed Bell diagonal states for $c_1 = 0.4, \; 0.8, \; 0.9, \; \text{and} \; 0.996$ whereas the $\blacksquare$ marker indicates the pure state entropy and the $\times$ marker indicates the Bell diagonal state entropy for $c_1=1.0$. Unlike in the SEAQT framework, the Bell diagonal states in the Lindblad framework for the choice of jump operators chosen are dissipative except for the case when $c_1 = c_2$ (see Eq.~(\ref{eq:eq_Lind})), which occurs for $c_1 = 0.4$. As seen in this figure,  the behavior in the Lindblad framework is mixed. For $c_1=0.996$ the total dissipation initially decreases and then monotonically increases as the initial perturbed state moves away from the Bell diagonal state whereas for $c_1=0.9$ it is always increasing. When $c_1=0.8$, there is both a monotonic increase and then decrease while for $c_1=0.4$ the total dissipation continually decreases as the initial perturbed state entropy moves closer to that of the pure state.  Furthermore, not only is the  final state in the Lindblad framework not consistent with the second law of thermodynamics but so is every non-stationary state along the path predicted by this equation and, thus, the dissipation predicted is not that which would result from the second law.  \\

\begin{figure} 	 [!htb]
		\includegraphics[scale=\FigScaleFact]{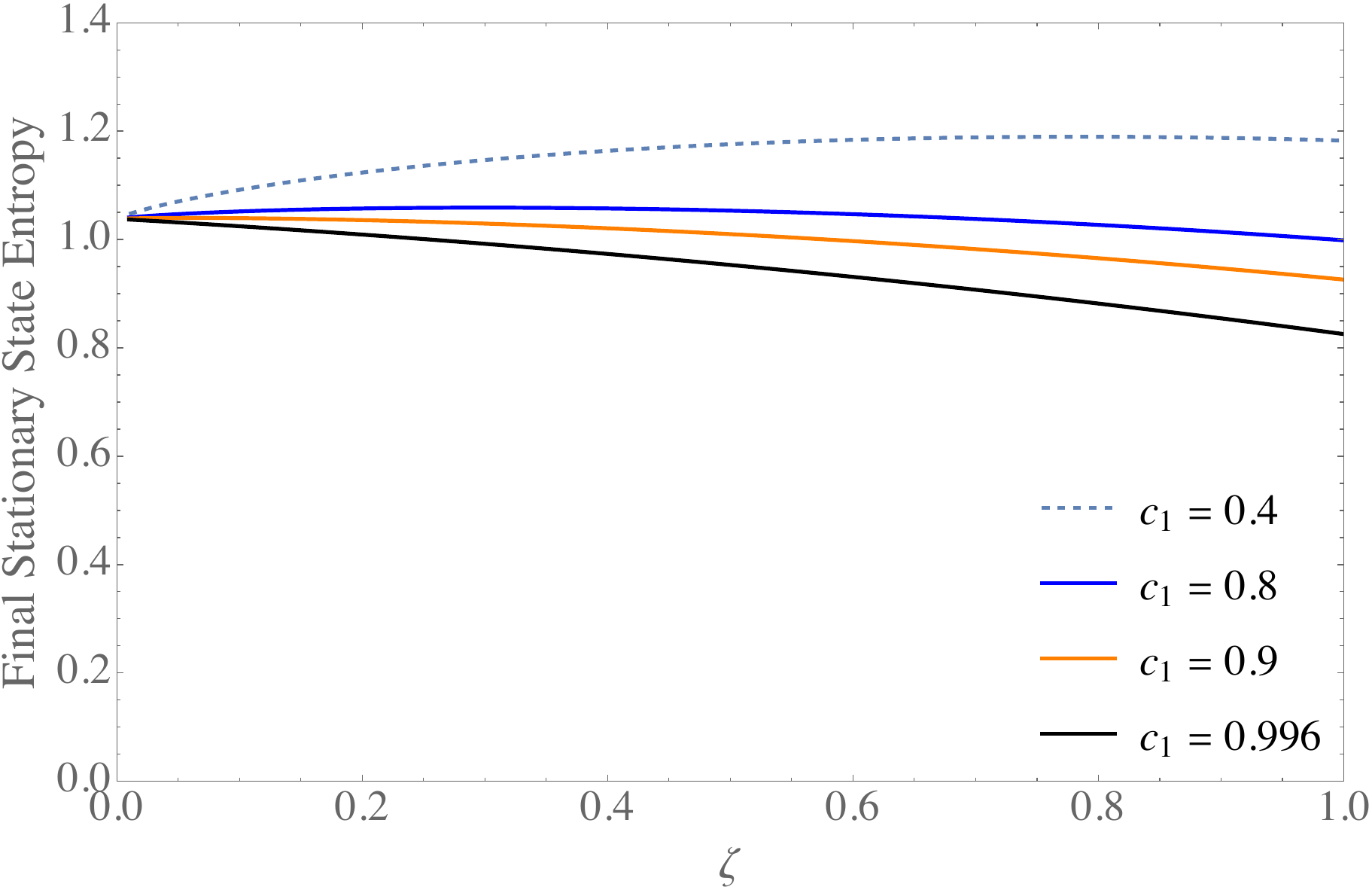} \\
	\caption{Dependence in the Lindblad framework between the weighting parameter $\zeta$ and the stationary state entropy for $c_1 = 0.1, 0.4, 0.8, 0.9$ and $0.996$.}
	\label{Fig:SS_EG_Lind}
\end{figure}

In Fig.~\ref{Fig:SS_EG_Lind}, both the stationary state entropy and the entropy generation as functions of the weighting parameter $\zeta$ are depicted. Notably, the highest stationary state entropies correspond to the perturbed states with $c_1 = 0.4$, while the entropy generation is the least for this case. Additionally, for $\zeta = 1$ and $c_1 = 0.4$, the entropy generation is zero, which, according to Eq.~(\ref{eq:eq_Lind}), implies the absence of entropy changes. This observation suggests a unique stationary (steady state) under these specific parameter conditions.

\subsection{Entanglement Evolution Initialized Via a General Bipartite Energy-Entropy Perturbation}
 \label{Sec:constrained}
\subsubsection{SEAQT framework} \label{Sec:constrainedSEAQT}
This section presents results using the general bipartite system perturbation approach with constant energy and constant linear entropy constraints and a variance of $\sigma=10^{-1}$. Specifically, results are presented for the evolutions of $E(\hat{\rho})$ and $\langle{\hat{B}_{CHSH}\rangle}_{\mx}$ as are results for the  dependence of the initial perturbed and final stable equilibrium state $E(\hat{\rho})$ and $\langle{\hat{B}_{CHSH}\rangle}_{\mx}$ and the total entropy generation and total entropy change.

\begin{figure} [!htb]
\centering
		a)\includegraphics[scale=\FigScaleFact]{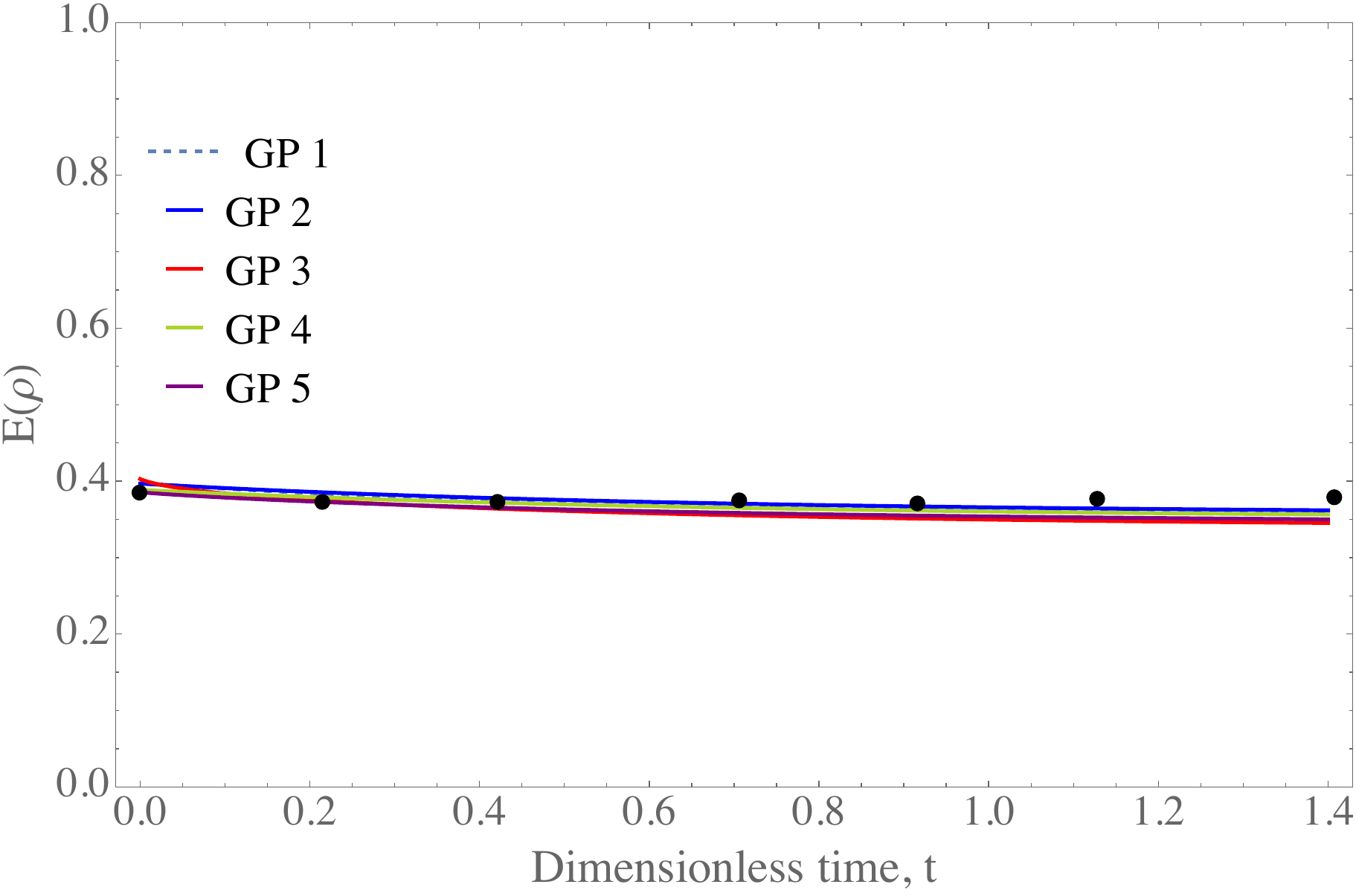} \\
		b) \includegraphics[scale=\FigScaleFact]{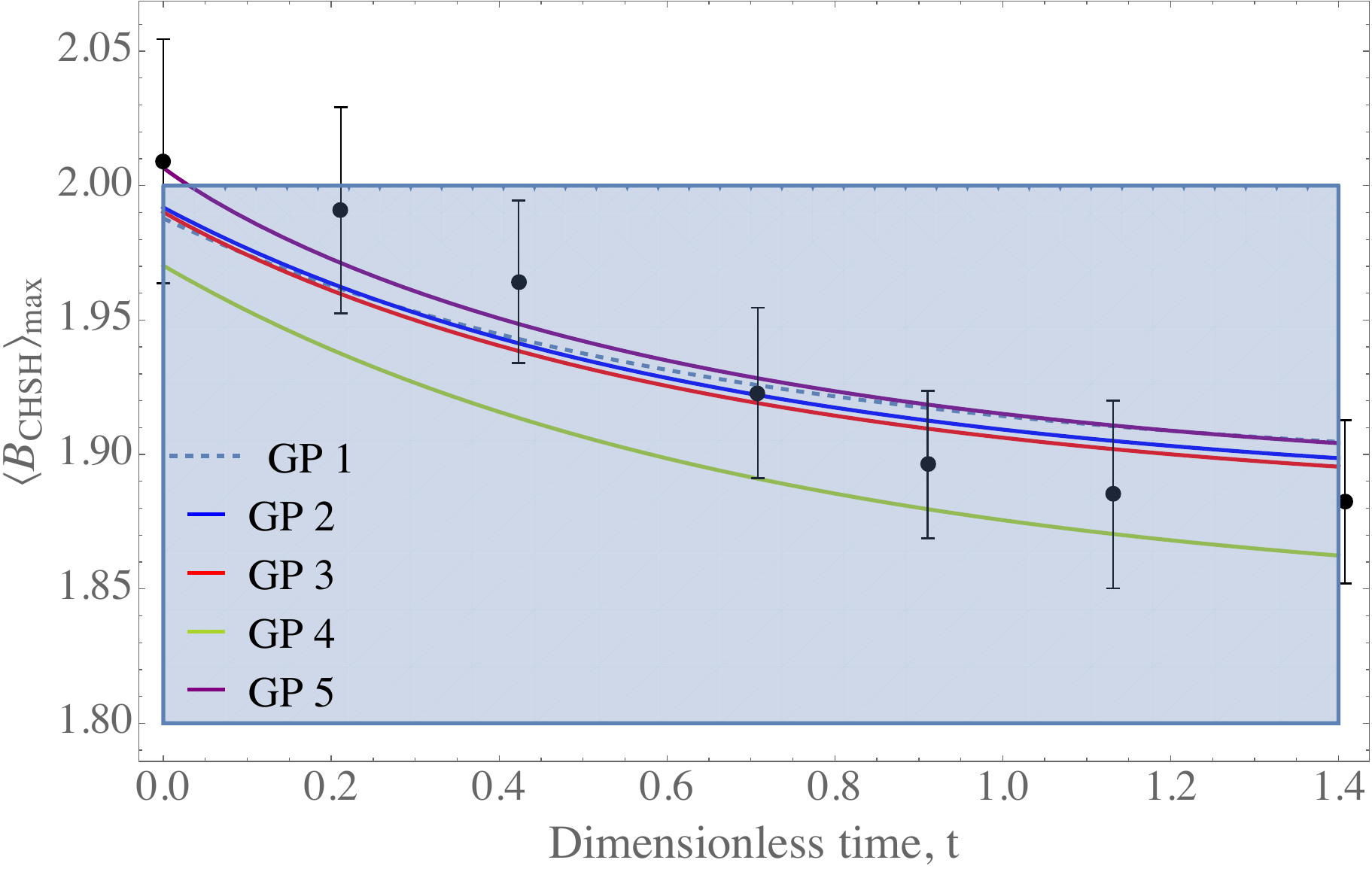}
			
	\caption{The a) $E(\hat{\rho})$ and b) $\langle{\hat{B}_{CHSH}\rangle}_{\mx}$ SEAQT evolutions for the five initial states generated with the general perturbation method.}
	\label{Fig:Cons_pert_SEAQT}
\end{figure}

Fig.~\ref{Fig:Cons_pert_SEAQT} shows the evolution within the SEAQT framework of $E(\hat{\rho})$ and $\langle{\hat{B}_{CHSH}\rangle}_{\mx}$ for five random perturbations. This choice of initial perturbed states is taken from 1,500 generated of which 11 are sufficiently close to the original Bell diagonal state of Lui $et \; al.$ \cite{liu_time-invariant_2016} so that both the SEAQT and Lindblad evolutions match the experimental data for the concurrence and $\langle{\hat{B}_{CHSH}\rangle}_{\mx}$. This is not the case for any of the evolutions associated with the weighted-average perturbations, i.e., either the experimental concurrence is matched or the $\langle \hat B_{CHSH} \rangle_{\mx}$ measure but not both. Of the 11 perturbations sufficiently close, the 5 chosen result in  the best fit of the experimental data as determined by the closeness of the initial perturbed states to the original Bell diagonal state. The closeness is determined using the relative entropy with GP1 to GP5 ordered from closest to the least close. Note that four of the five evolutions maintain a value of $\langle \hat B_{CHSH} \rangle_{\mx}$ below the non-locality condition at all times. Of the 1,500 initial perturbed states generated, a number do exhibit non-local entanglement. Furthermore,  these evolutions of the concurrence as well as of the $\langle \hat B_{CHSH} \rangle_{\mx}$ measure will decay further since stable equilibrium has not been reached for the final values shown in the figure.

\begin{figure}  [!htb]
\centering
	a)\includegraphics[scale=\FigScaleFact]{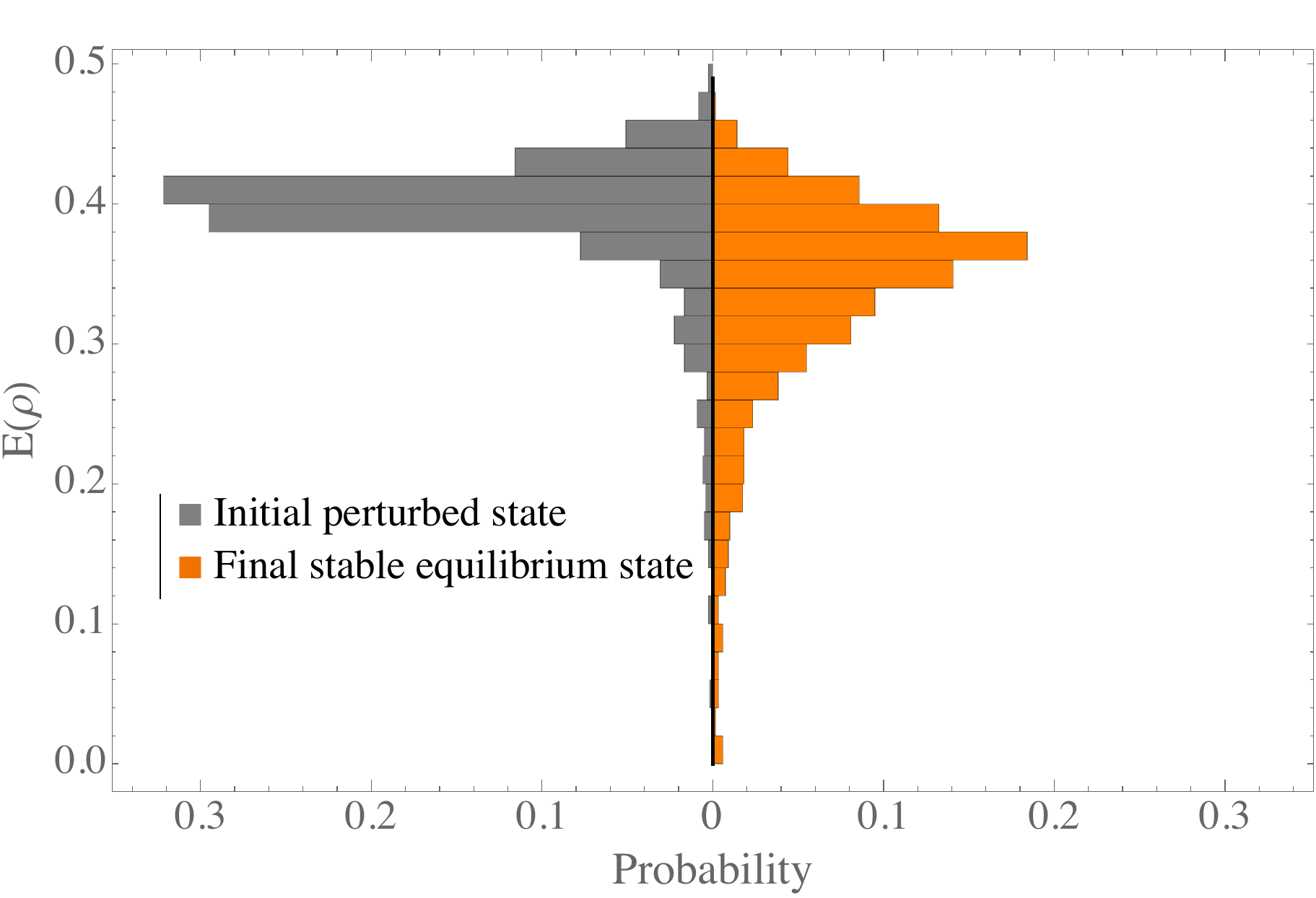}
	b)\includegraphics[scale=\FigScaleFact]{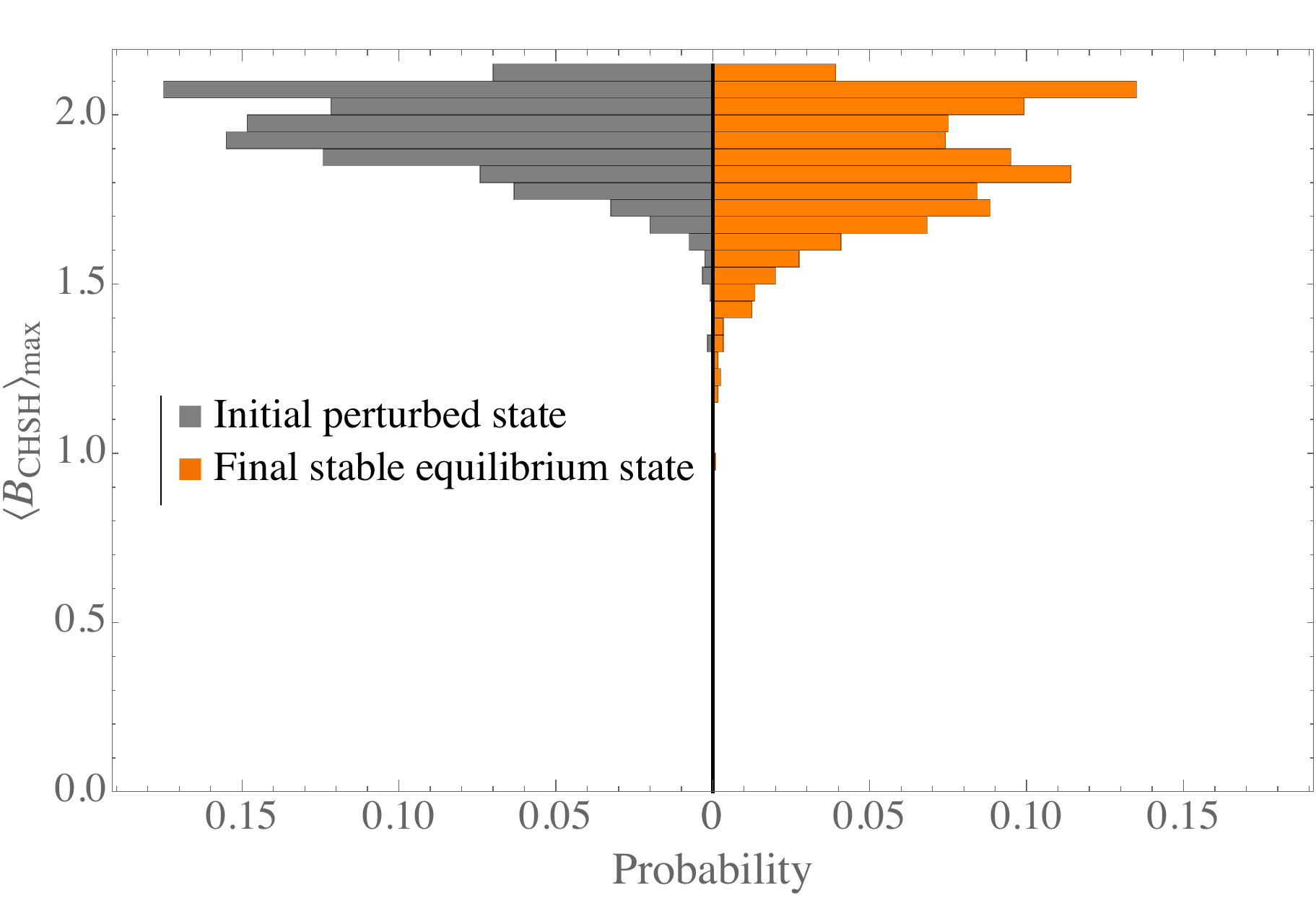}
	\caption{Initial perturbed state and final stable equilibrium state histograms for a) $E(\hat{\rho})$ and b) $\langle{\hat{B}_{CHSH}\rangle}_{\mx}$ in the SEAQT framework. The gray bars correspond to the initial state values whereas the orange bars to the final state values. The mean value for the initial perturbed states is 1.94, whereas that for the final state is 1.85 for $\langle\hat{B}_{CHSH}\rangle_{\text{max}}$. In the case of $E( \hat \rho )$, the mean value of the initial perturbed states is 0.39 and for the final stable equilibrium state, 0.34.} 
	\label{Fig:hist_SEAQT}
\end{figure}

Fig. \ref{Fig:hist_SEAQT} a) shows histograms of the concurrence for the initial and final states obtained from 1,500 random perturbations with $\eta \in \text{GUE}$ and mean and standard deviations of 0 and $10^{-1}$, respectively. It is observed that the initial concurrence has a mean value of 0.38, closely matching the concurrence of the pure Bell diagonal state. This is expected, as the constrained perturbation assures that the initial perturbed state has a concurrence similar to that of the original Bell diagonal state. Moreover, at the end of the dynamics, the average concurrence has decreased somewhat, indicating a reduction in entanglement. In contrast, Fig. \ref{Fig:hist_SEAQT} b) shows that the $\langle B_{\text{CHSH}} \rangle_{\mx}$ exhibits higher dispersion for both the initial and final states. Furthermore, considering the limit of non-locality in states where $\langle B_{\text{CHSH}} \rangle_{\mx} \leq 2$, it is found that the random perturbations result in nearly 32 \% of states exhibiting non-locality properties at the end of the dynamics.

\begin{figure} [!htb]		
	\centering
		a) \includegraphics[scale=\FigScaleFact]{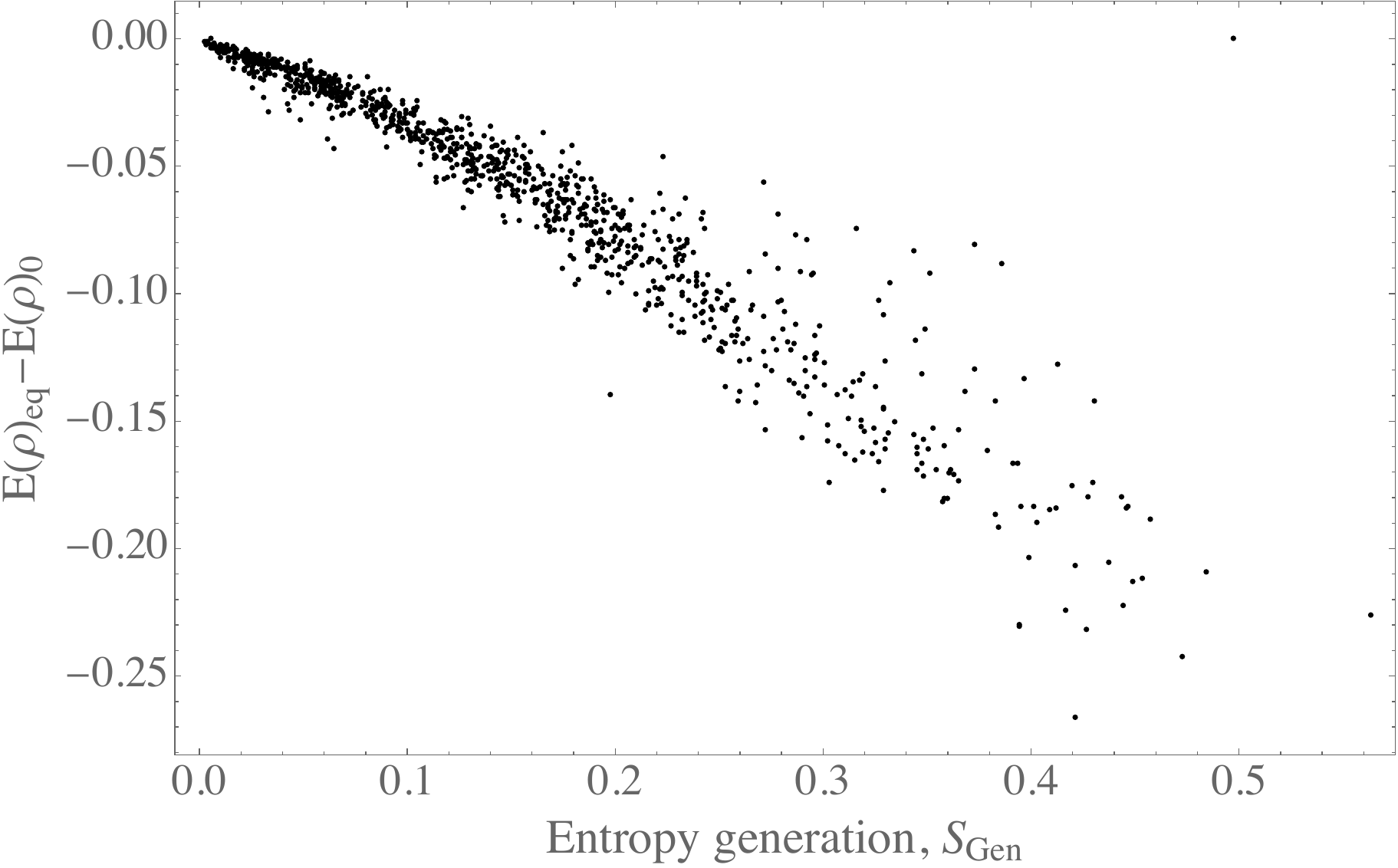}\\
		b) \includegraphics[scale=\FigScaleFact]{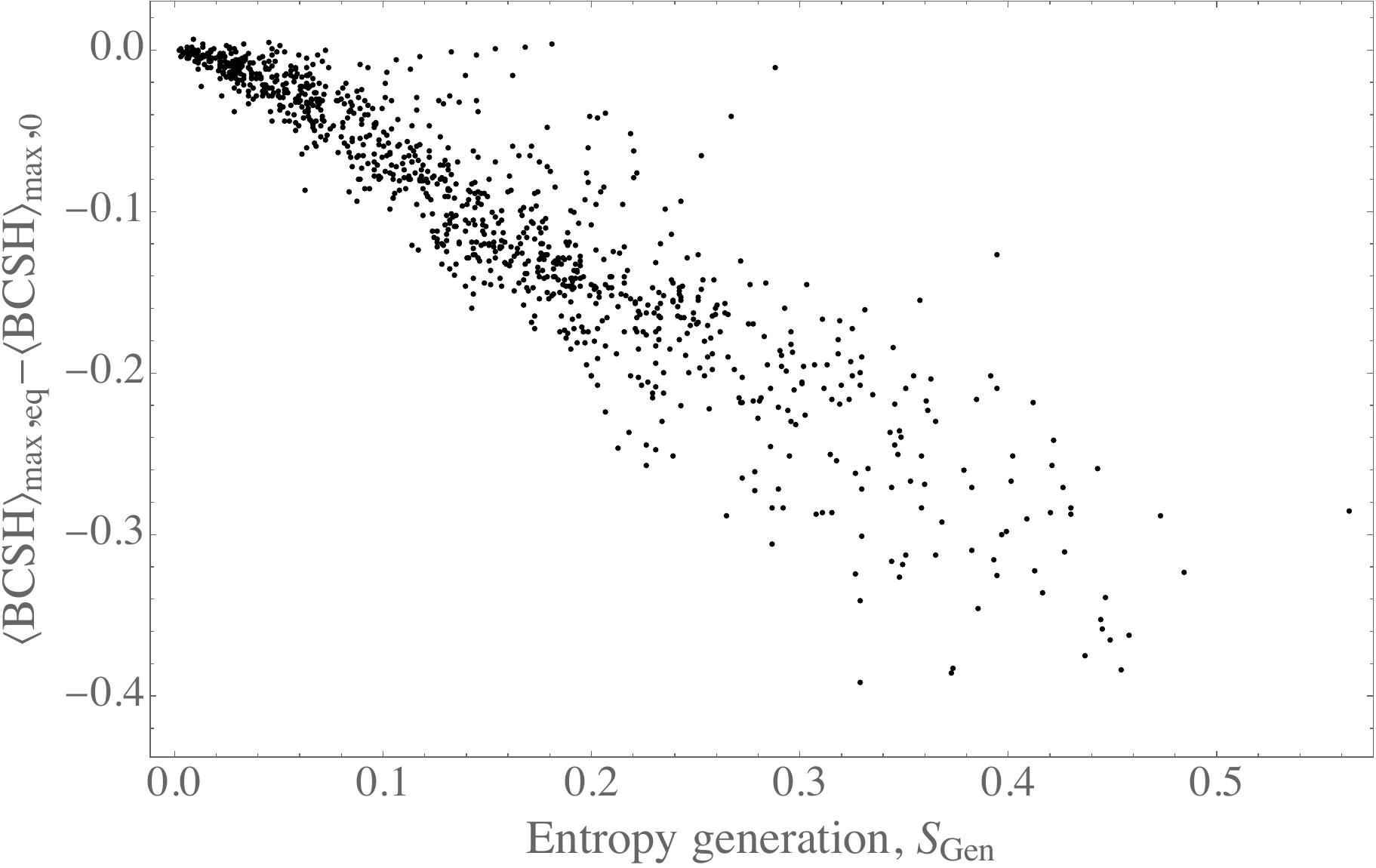}
	\caption{Dependence of  a) $E(\hat{\rho})$ and b) $\langle{\hat{B}_{CHSH}\rangle}_{\mx}$ on the total entropy generation in the SEAQT framework.}
	\label{Fig:deltaEB_Sgen_SEAQT}
\end{figure}

Fig. \ref{Fig:deltaEB_Sgen_SEAQT} shows the dependence on the total entropy generation (or production) of the total change of $E(\hat{\rho})$ and $\langle{\hat{B}_{CHSH}\rangle}_{\mx}$ from the initial perturbed state to the final stable equilibrium state. Both properties show clear trends of decrease with increasing entropy generation. One interesting aspect to note is that while the average $E(\hat{\rho})$ is seen to only decrease relative to the SEAQT dynamics, for a few of the initial perturbed states the evolution resulting from this dynamics does cause an increase in $\langle{\hat{B}_{CHSH}\rangle}_{\mx}$. Since the SEAQT equation of motion and the resulting dissipation do not increase the entanglement for two non-interacting subsystems, this increase indicates that these few states are sufficiently close to the Bell diagonal state so that there is no decay in entanglement or non-locality. Thus, for non-monotonic measures, like $\langle{\hat{B}_{CHSH}\rangle}_{\mx}$, a less entangled state can have a higher value of $\langle{\hat{B}_{CHSH}\rangle}_{\mx}$ than a more entangled state (e.g., a completely unentangled state can have a value as high as 2 whereas entangled states can have values less than or greater than 2).

Additionally,  Fig. \ref{Fig:deltaEB_Sgen_SEAQT} illustrates that an increase in the information measure corresponds to an increase in entropy generation. This dependence can be measured by considering the entropy generation as well as the decay of $E (\hat \rho)$ or $\langle B_{\text{CHSH}}\rangle_{\mx}$ as random variables with Pearson coefficient
\begin{equation}
r_{E(B)} = \frac{\text{Cov}(\delta E(\hat{\rho}), S_{\text{Gen}})}{\sqrt{\text{Var}(\delta E(\hat{\rho})) \cdot \text{Var}(S_{\text{Gen}})}} 
\end{equation}
where $\delta E(\hat{\rho} ) = E(\hat{\rho})_\text{eq} - E(\hat{\rho})_0 $ and similarly for $\langle B_{\text{CHSH}}\rangle_{\mx}$. In this case, the Pearson coefficients are $r_{E} \approx -0.94$ and $r_B \approx -0.91$ indicating a strong negative correlation between the information measures and the entropy generation, i.e., that as the entropy generation increases the change in $E(\hat \rho )$ as well as $\langle B_{\text{CHSH}}\rangle_{\mx}$ reduces.

\subsubsection{Lindblad framework}
\label{Sec:SEAQT_constrained}
\begin{figure} [!htb]
\centering
		a)\includegraphics[scale=\FigScaleFact]{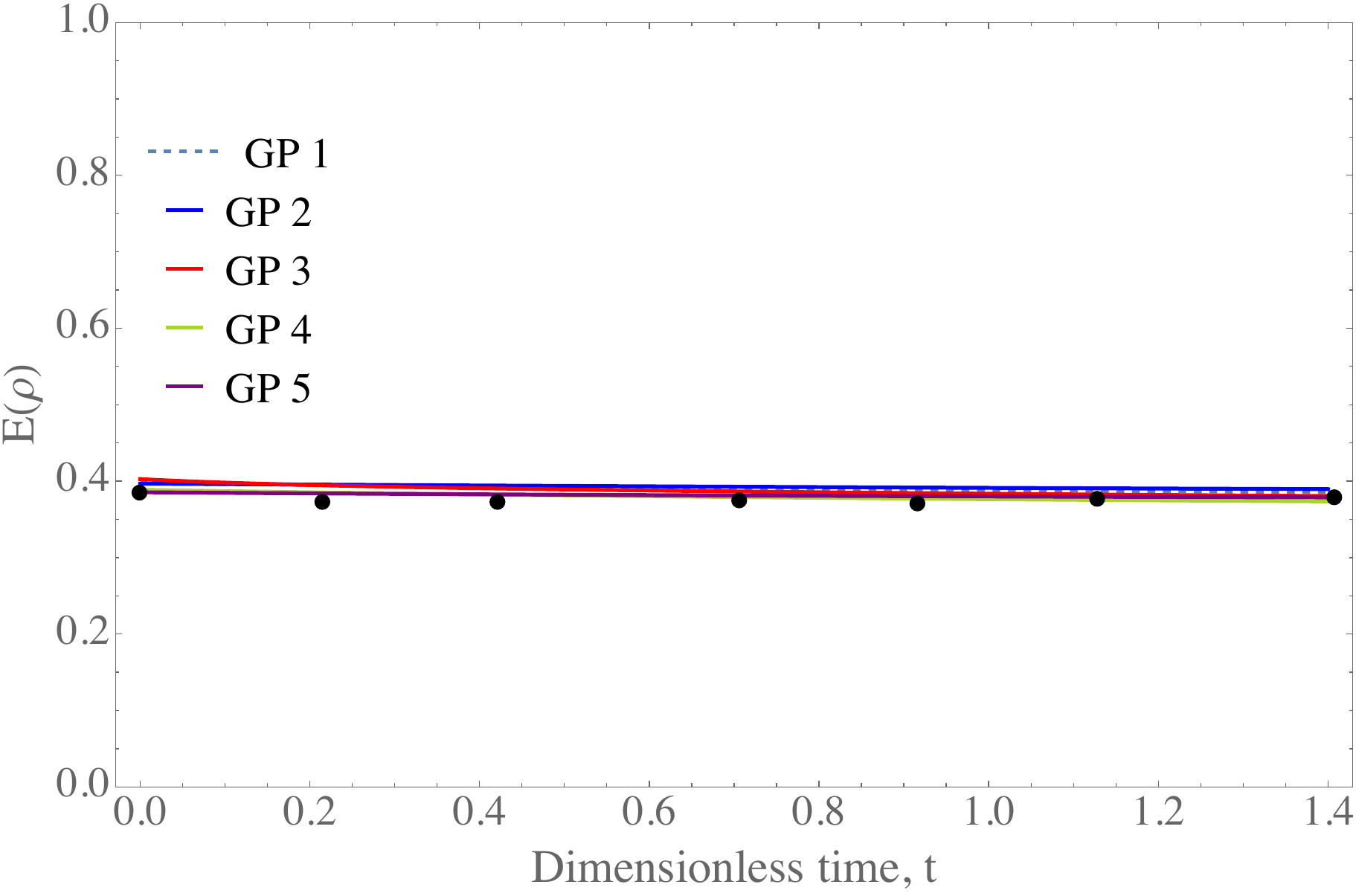} \\
		b) \includegraphics[scale=\FigScaleFact]{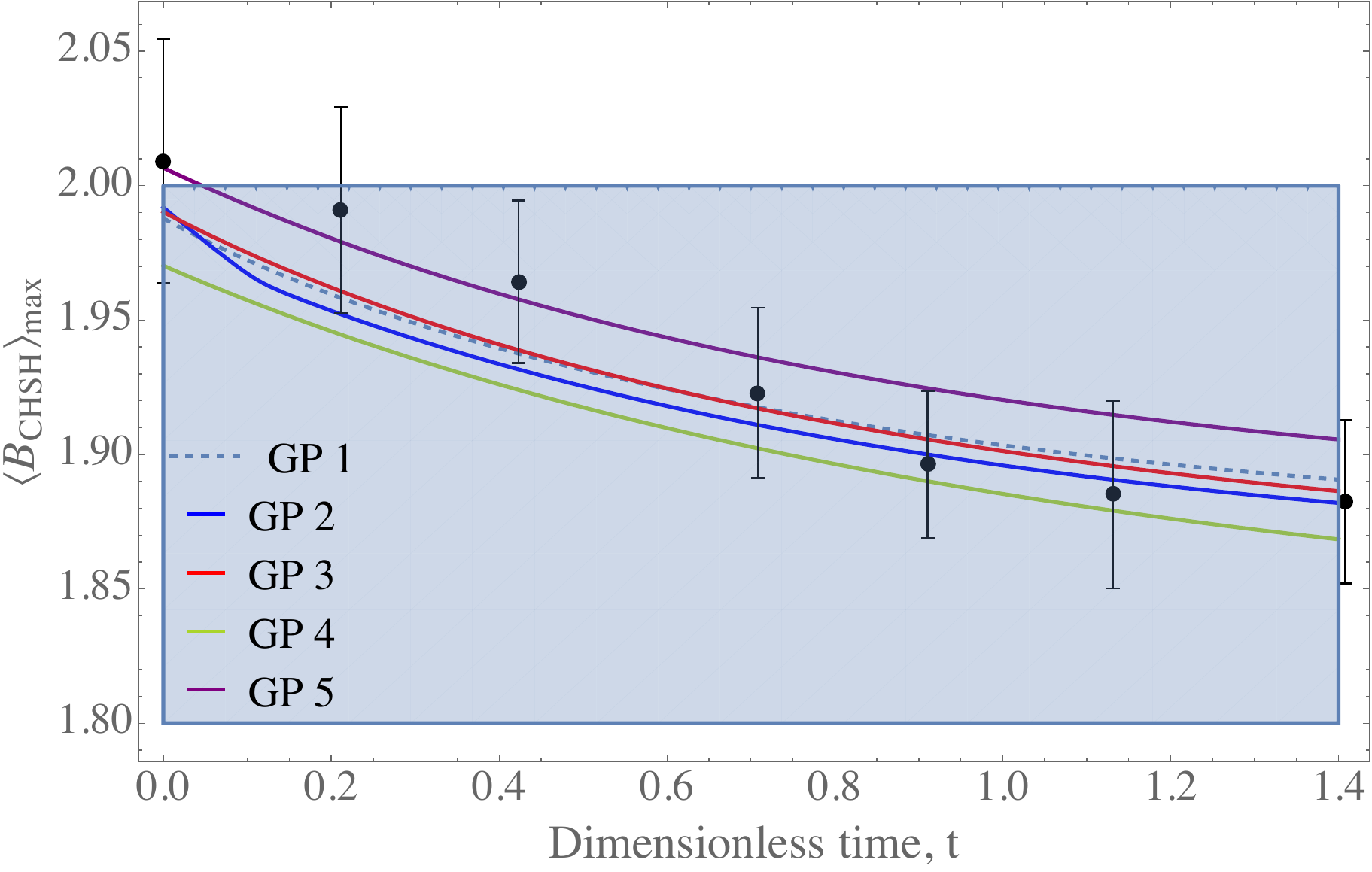}
			
	\caption{a) $E(\hat{\rho})$ and b) $\langle{\hat{B}_{CHSH}\rangle}_{\mx}$ evolutions based on the Lindblad framework for the five random initial states generated using the general perturbation method.}
	\label{Fig:Cons_pert_Lind}
\end{figure}

Results for the evolutions of $E( \hat \rho )$ and $\langle \hat B_{CHSH} \rangle_{\mx}$ based on the Lindblad framework are given in this section as is the dependence of $E( \hat \rho )$ and $\langle \hat B_{CHSH} \rangle_{\mx}$ on the stationary states found using the Lindblad framework and the same 5 general perturbation initial states used in the SEAQT evolutions. As observed in Fig. \ref{Fig:Cons_pert_Lind}, the evolution of the concurrence for the five random perturbations shows some decay but remains close to the 0.4 value reported in the experiments, whereas for $\langle \hat B_{CHSH} \rangle_{\mx}$, the decay in time is more pronounced. Four of the five evolutions predict a death of non-locality at all values of the dimensionless time. Of the 1,500 initial perturbed states generated, a number, as noted before, do exhibit non-local entanglement.

\begin{figure}  [!htb]
\centering
	a)\includegraphics[scale=\FigScaleFact]{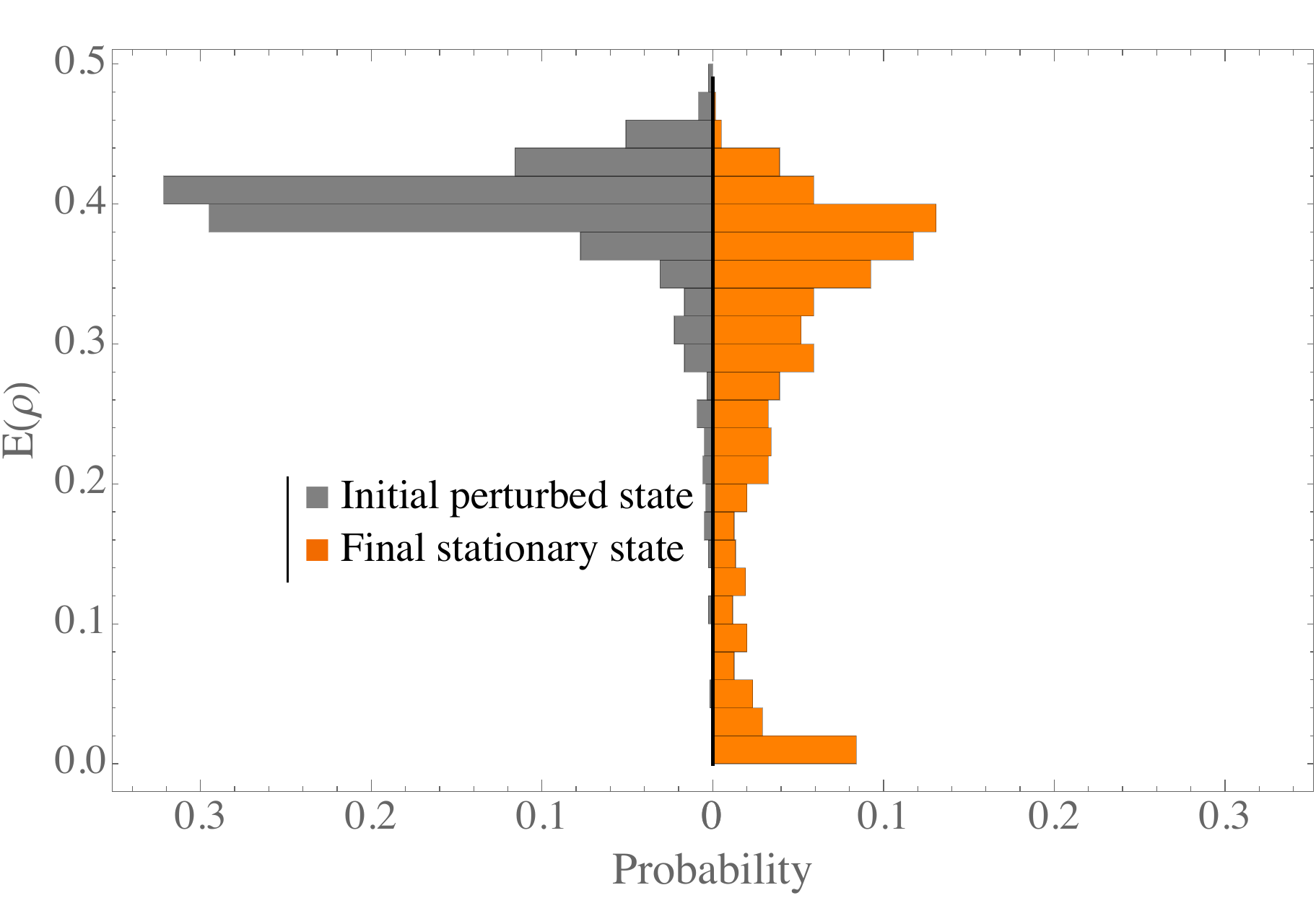}
	b)\includegraphics[scale=\FigScaleFact]{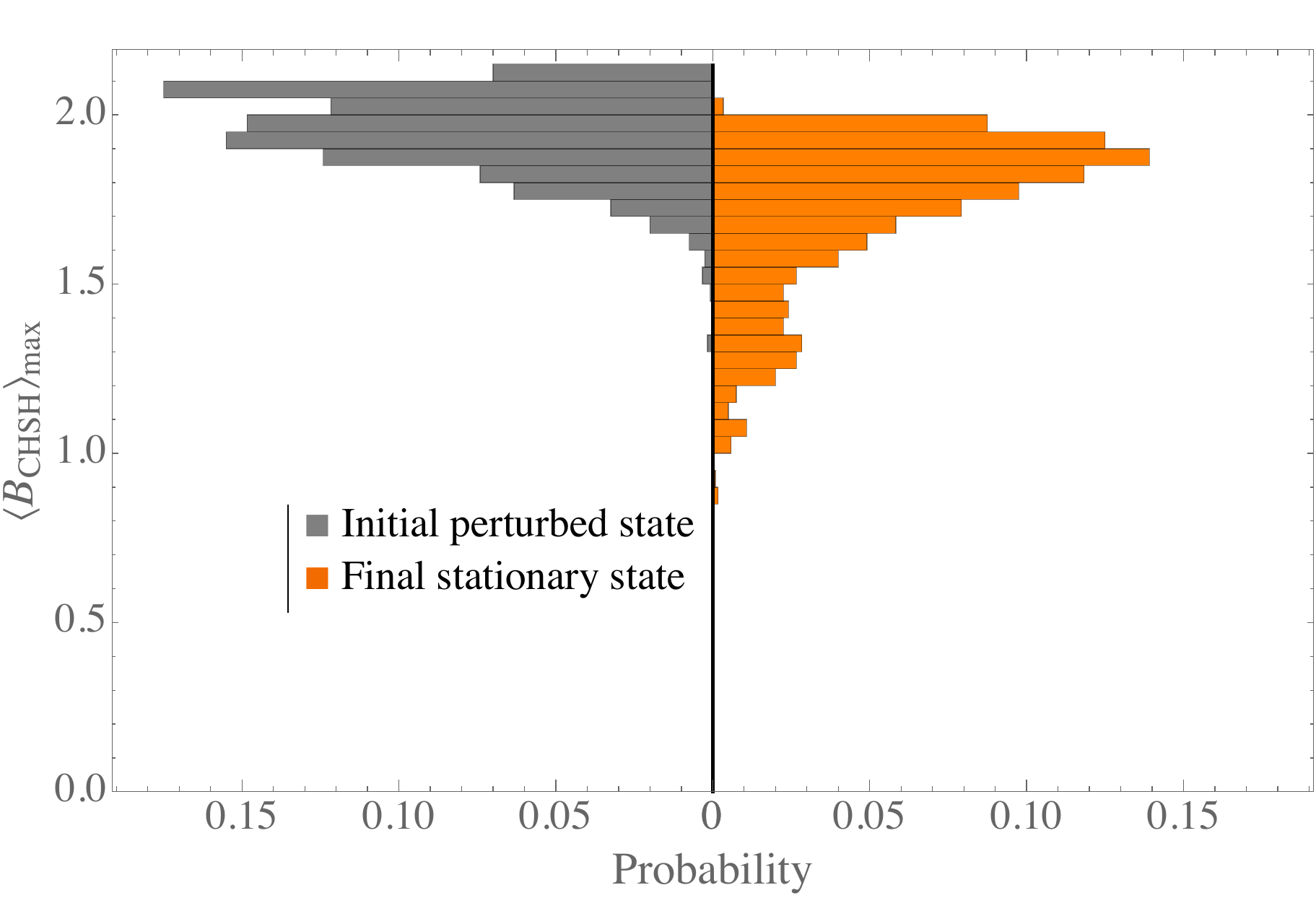}
	\caption{Initial perturbed state and final stationary state histograms for a) $E(\hat{\rho})$ and b) $\langle{\hat{B}_{CHSH}\rangle}_{\mx}$ in the Lindblad framework. The gray bars correspond to the initial state values and the orange bars to the final state values. The mean value for the initial perturbed states is 1.94, whereas that for the final state is 1.71 for $\langle\hat{B}_{CHSH}\rangle_{\text{max}}$. In the case of $E( \hat \rho )$, the mean value of the initial perturbed states is 0.39 and for the final stable equilibrium state,  0.27.}
	\label{Fig:hist_Lind}
\end{figure}

Fig.~\ref{Fig:hist_Lind} presents histograms based on 1,500 constrained state perturbations generated using the general perturbation method and the corresponding stationary states found using the Lindblad dynamics. In Fig.~\ref{Fig:hist_Lind} a), the mean concurrence value for the initial perturbed states (gray bars) is approximately 0.4. This is a consequence of selecting specific roots from the set of constraints for initializing the density matrix close to the concurrence of the reference Bell diagonal state.  Under the Lindblad dynamics, the average concurrence tends to decrease, exhibiting greater variance than that predicted by the SEAQT dynamics. Conversely, as shown in Fig.~\ref{Fig:hist_Lind} b), the initial and final state values of the $\langle \hat B_{CHSH} \rangle_{\mx}$ measure display a less broad dispersion compared to the concurrence histograms. Furthermore, the value of the $\langle \hat B_{CHSH} \rangle_{\mx}$ measure is less than 2.0 for 75 \% of the initial perturbed states, while at the end of the dynamics, all the final states have a value of this measure less than 2.0, indicating a loss of non-locality in all cases.

\begin{figure} [!htb]		
	\centering
		a) \includegraphics[scale=\FigScaleFact]{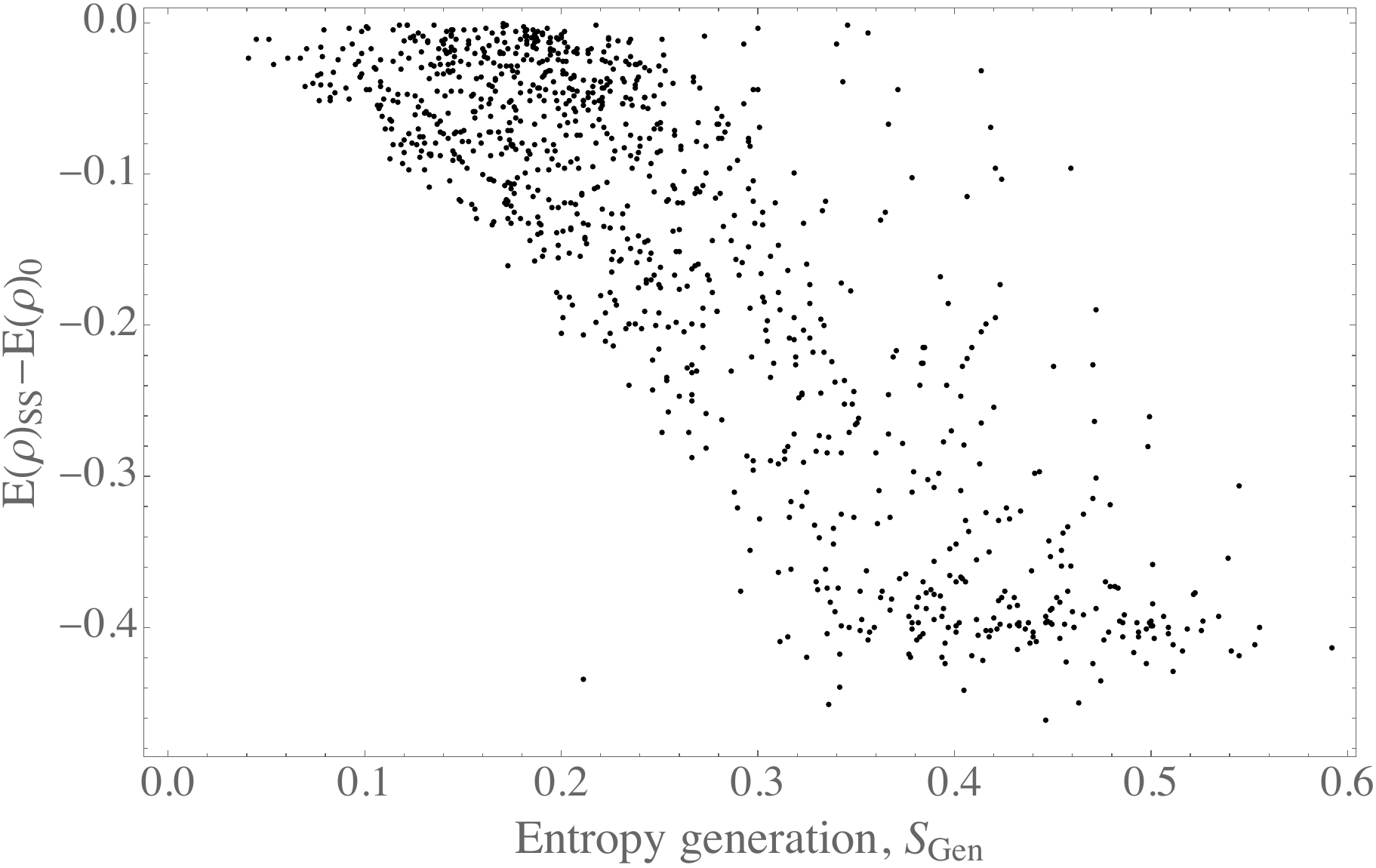}\\
		b) \includegraphics[scale=\FigScaleFact]{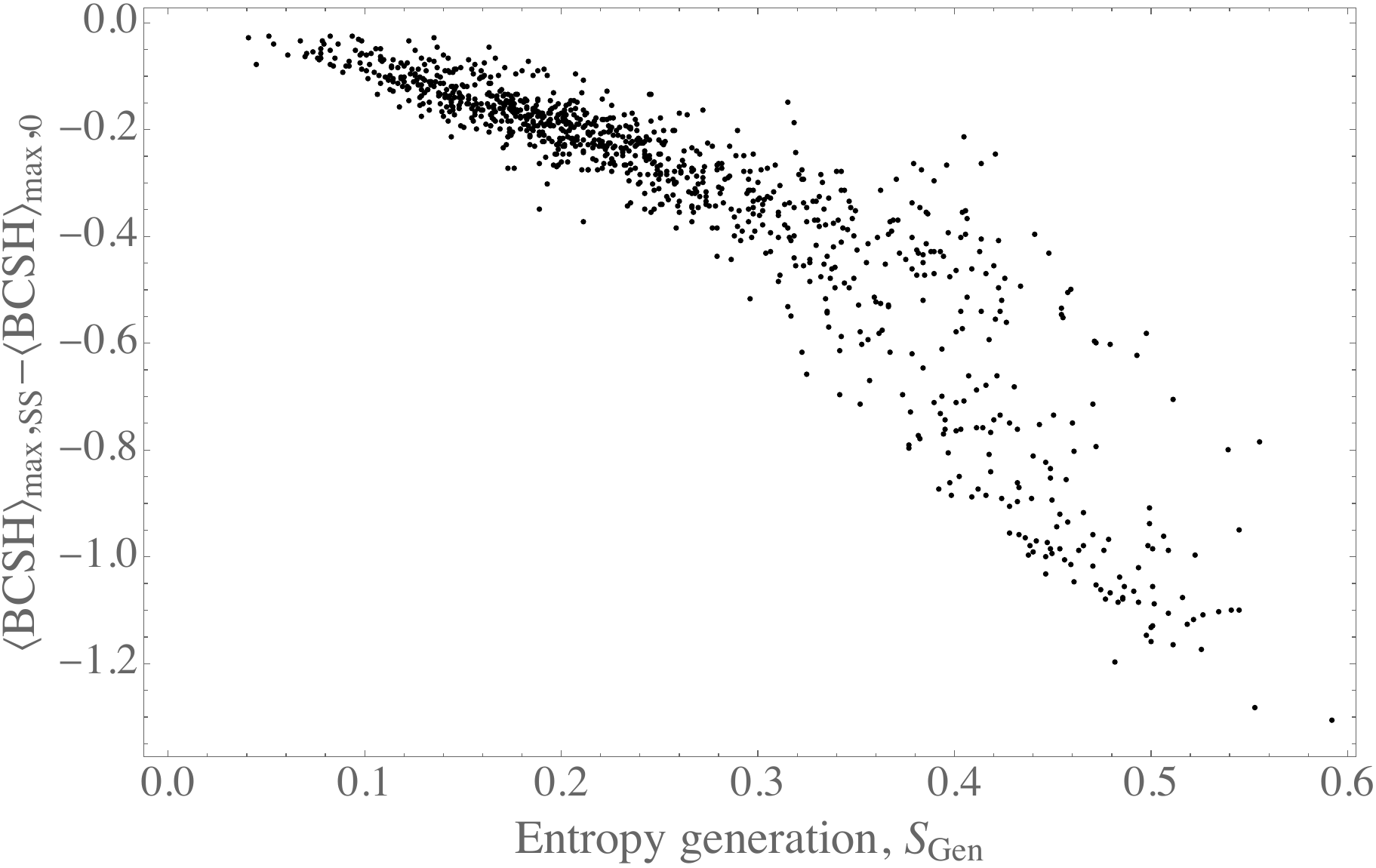}
	\caption{Dependence of a) $E(\hat{\rho})$ and b) $\langle{\hat{B}_{CHSH}\rangle}_{\mx}$ on the total entropy generation in the Lindblad framework.}
	\label{Fig:deltaEB_Sgen_Lind}
\end{figure}

Finally, the scatter plots in Figure~\ref{Fig:deltaEB_Sgen_Lind} depict the correlations of the changes in the concurrence and the $\langle \hat B_{CHSH} \rangle_{\mx}$ measure relative to the entropy generation with Pearson coefficients of $r_E \approx -0.83$ and $r_B \approx -0.91$. The former indicates a weaker correlation between the change in the concurrence and the entropy generation than that exhibited with the SEAQT dynamics, while the latter indicates that the correlation between the change of the $\langle \hat B_{CHSH} \rangle_{\mx}$ measure and the entropy generation is almost the same as that in the SEAQT dynamics. Notably, as the entropy generation increases, the spread in the concurrence and the $\langle \hat B_{CHSH} \rangle_{\mx}$ measure values increases. This is true for the SEAQT dynamics as well although the increase in the spread is not as pronounced.

 Note that although the perturbations of only five initial states generated by the general bi-partite perturbation scheme are considered here, many more such states were generated (1,500, in fact). As seen in the work of Monta\~nez $et \; al.$ \cite{montanez-barrera_method_2022}, the general bi-partite perturbation scheme accurately predicts large numbers of states in the vicinity of those generated by a real quantum computer. The authors demonstrate that both the entropy and energy constraints play an important role in generating states comparable to those of a real device where noise prevents the device itself from generating ideal states, i.e., Bell or Bell diagonal states.

\subsection{Comparison between the SEAQT and Lindblad frameworks}
\begin{figure} [!htb]		
	\centering
		a) \includegraphics[scale=\FigScaleFact]{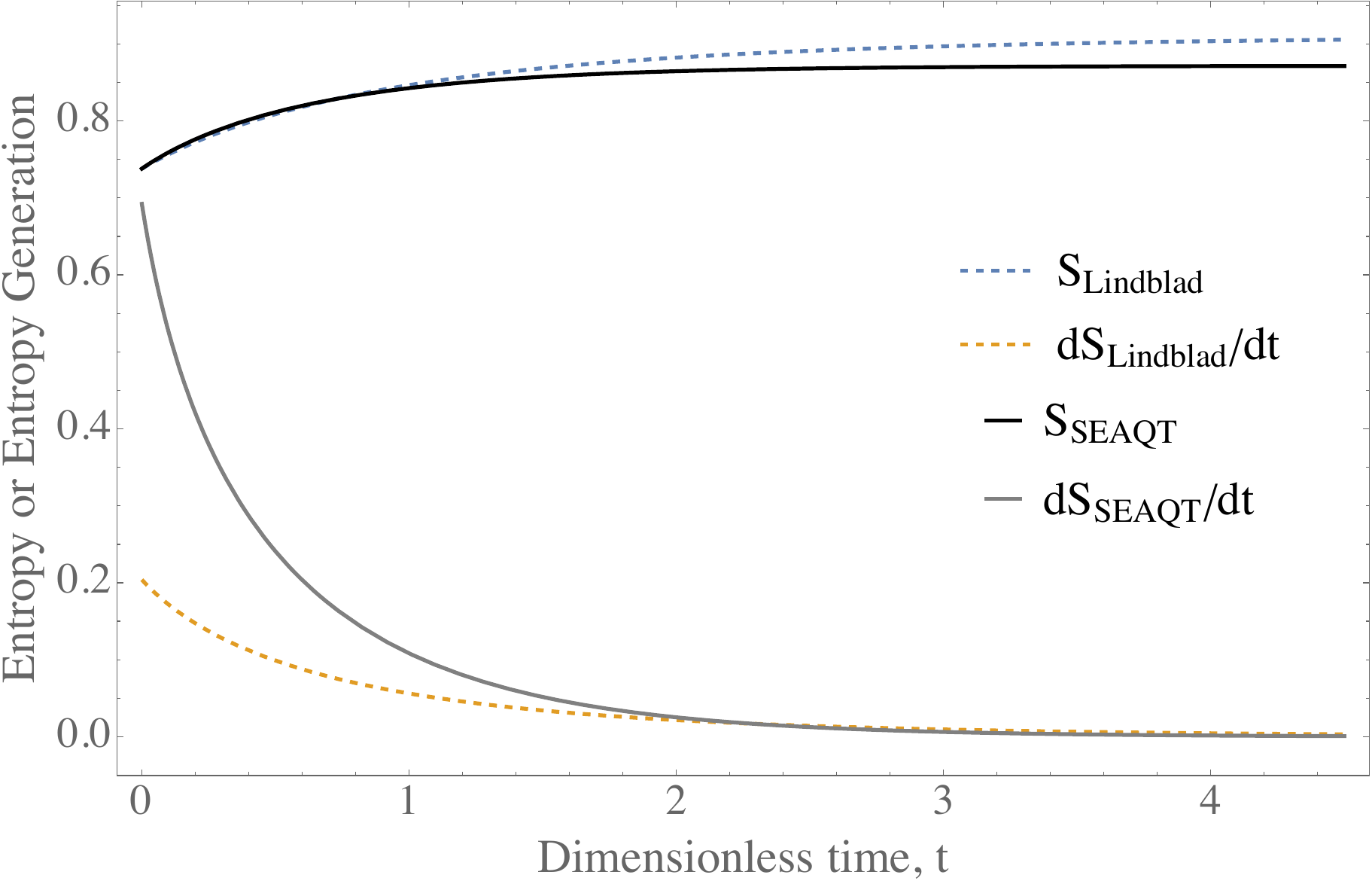}\\
		b) \includegraphics[scale=\FigScaleFact]{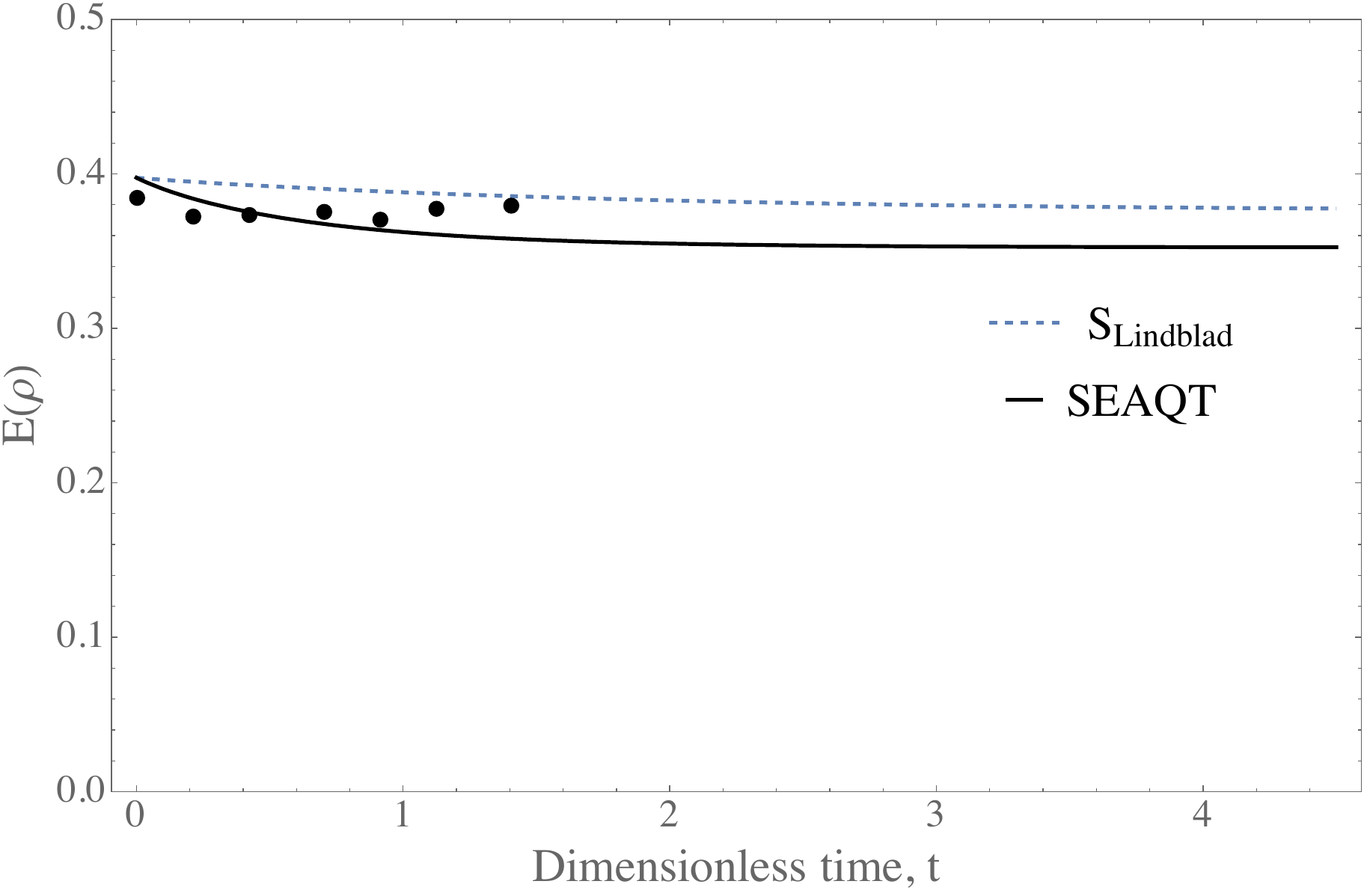}\\
		c) \includegraphics[scale=\FigScaleFact]{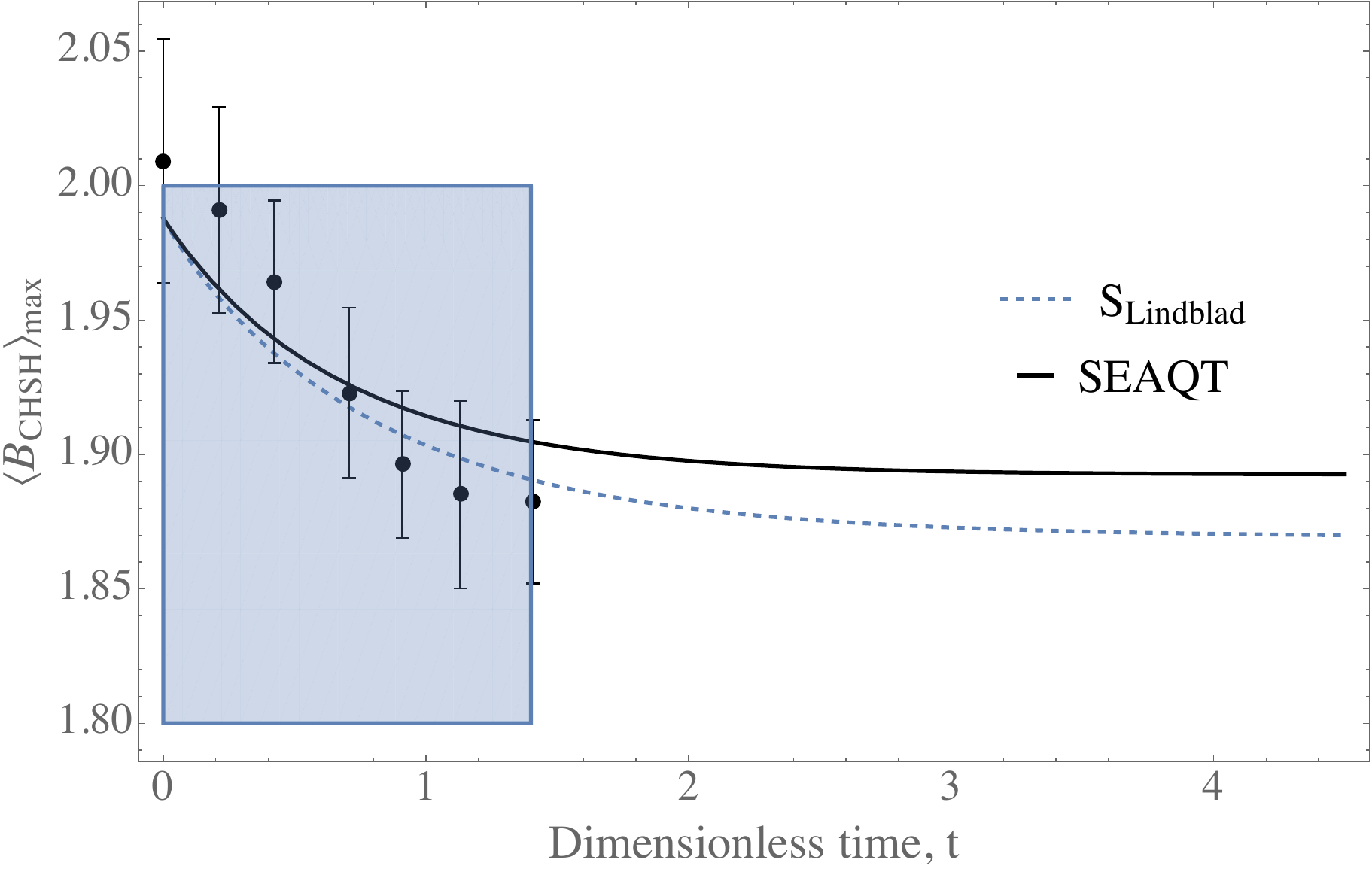}
	\caption{Evolution of a) the entropy and the time rate of change of the entropy, b) the $E(\rho)$ and c) the $\langle \hat B_{CHSH} \rangle_{\mx}$ measure in the SEAQT and Lindblad frameworks. The experimental data is include in figures b) and c).}
	\label{Fig:SEAT_Lind}
\end{figure}

Fig.~\ref{Fig:SEAT_Lind} illustrates the dynamics of the entropy, $E(\hat \rho)$, and $\langle \hat B_{CHSH} \rangle_{\mx}$ in both the SEAQT and Lindblad frameworks. The initial perturbed state used for both the SEAQT and Lindblad evolutions is GP1. 
As seen in Fig.~\ref{Fig:SEAT_Lind} a), the Lindblad approach consistently results in a higher entropy than that predicted by the SEAQT equation of motion at each instant of time. In contrast, the time rate of change of the entropy is lower in the Lindblad framework, indicating a less rapid approach to the final stationary state in this framework than to the final stable equilibrium state in the SEAQT framework. As to the concurrence, Fig.~\ref{Fig:SEAT_Lind} b) shows that both models exhibit a relatively stable value over time with only slight fluctuations. Intriguingly, in the context of the $\langle \hat B_{CHSH} \rangle_{\mx}$ measure, Fig.~\ref{Fig:SEAT_Lind} c) shows that both frameworks concur on the absence of non-locality, a finding borne out by the experimental data. This lack of non-locality in both frameworks highlights the robustness of this phenomenon across different theoretical approaches and reinforces its empirical validity.

\section{Conclusions}
\label{Sec:Conclusions}

This paper has presented a study of the evolution of a bipartite system in an entangled state using the dynamics of the SEAQT and Lindblad frameworks and provided insight into the entanglement dynamics within both frameworks for non-interacting subsystems. Results for these evolutions are indirectly compared to evolutions predicted by the Kraus-Operator-Sum approach as well as to data obtained from a quantum optics experiment. While the dynamics predicted by the SEAQT framework differ from those predicted by the Kraus-Operator-Sum approach, both predict the possibility of almost constant concurrence and the presence of quantum entanglement or non-locality (via the CHSH operator maximum expectation value being above 2). In addition, SEAQT and Lindblad frameworks predict the sudden death of non-locality via a drop below 2 of the maximum CHSH operator expectation value at dimensionless times consistent with the experimental data but much smaller than the times predicted by the Kraus-Operator-Sum approach. 

Using the weighted-perturbation approach to generate the initial states, a good match between the experimental bounds and the predictions made with the SEAQT and Lindblad frameworks for the behavior of the $\langle \hat B_{CHSH} \rangle_{\mx}$ measure is found. At the same time, a significant discrepancy with respect to $E ( \hat{\rho})$ occurs. In contrast, using the general perturbation approach to generate the initiate states, predictions by the dynamics of both the SEAQT and Lindblad frameworks of the behavior of $\langle \hat B_{CHSH} \rangle_{\mx}$ and $E (\hat{\rho})$ match quite well with the experimental results. These results furthermore suggest that the experimental sets are not able to create a diagonal state, but a state close to it with the same energy and entropy. 

Finally, both information measures show a strong negative correlation, represented by Pearson coefficients less than -0.8, with the entropy generation, indicating the deep connection between thermodynamics and quantum information.  Remarkably, the SEAQT shows a higher correlation coefficient for $E(\hat{\rho})$ than the Lindblad formulation. This could be related to the fact that the SEAQT framework admits of a fluctuation-dissipation formulation, which relates the entropy generation to the covariance (fluctuation) of the generalized Massieu function. This stronger correlation is also based on the fact that the entropy generated in this unified framework of quantum mechanics and thermodynamics is based on a principle intimately tied to the second law of thermodynamics. It is this principle that leads to the non-linear dynamics of the SEAQT equation of motion and that captures the energy transition rates that take place between the discrete energy levels of the system. It is the contention of this unified theory that these transition rates driven by the principle of SEA are the underlying basis for all phenomena whether at an atomistic, mesoscopic, or macroscopic level. An equivalent principle intimately connecting the Lindblad (or equivalent Kraus-Operator-Sum) formulation to thermodynamics is not present, and, thus, generally, this model does not evolve to a final stationary state that is equivalent to the canonical or Gibbs state of equilibrium thermodynamics. Of course, the opposite is true for the SEAQT framework. Further investigation of this intriguing connection to thermodynamics of the decay of these information measures using the SEAQT framework is left for future work.

\section{Acknowledgment}
C. Damian and A. Saldana would like to thank the Mexican National Council on Science and Technology for their financial support. C. Damian thanks the Universidad de Guanajuato for their support through the project CIIC 039/2023. C. Damian and A. Saldana thank M. R. von Spakovsky and Virginia Tech for their kind hospitality in the development of this project. Finally, R. T. Holladay would like to thank the U.S. Department of Defense for its support via the Science, Mathematics, and Research for Transformation (SMART) scholarship.



%

\end{document}